\documentclass[aps,pre,reprint]{revtex4-1}
\usepackage{graphicx,latexsym}
\usepackage{amsmath}
\usepackage{epstopdf}
\usepackage{color}
\usepackage{subfigure}
\usepackage{wasysym}

\def\diff{\mathop{\rm\mathstrut d\!}\nolimits}

\begin{document}

\title{Integral equation study of soft-repulsive dimeric fluids}

\author{Gianmarco Muna\`o$^{1,*}$,Franz Saija$^{2,\dag}$}
\affiliation{
$^{1}$Dipartimento di Scienze Matematiche e Informatiche, Scienze Fisiche 
e Scienze della Terra,
Universit\`a degli Studi di Messina,
Viale F.~Stagno d'Alcontres 31, 98166 Messina, Italy. \\}
\affiliation{
$^{2}$CNR-IPCF, Viale F.~Stagno d'Alcontres 37, 98158 Messina, Italy.}
\thanks{Corresponding author: gmunao@unime.it \\
$^{\dag}$franz.saija@cnr.it}
%\thanks{$^2$franz.saija@cnr.it}

\begin{abstract}
We study fluid structure and water-like anomalies of a system 
constituted by  dimeric
particles interacting via a purely repulsive 
core-softened potential by means of integral 
equation theories. In our model, dimers interact
through a repulsive pair potential of inverse-power form with a softened 
repulsion strength. By employing the Ornstein-Zernike approach and the 
reference interaction site model (RISM)
theory, we study the behavior of water-like anomalies upon progressively
increasing the elongation $\lambda$ of the dimers from the monomeric case 
($\lambda=0$) to the tangent configuration ($\lambda=1$). For each value of
the elongation we consider two different values of the interaction potential,
corresponding to one and two length scales, with the aim to provide a 
comprehensive description of the possible fluid scenarios of this model.
Our theoretical results are systematically compared with already existing
or newly generated Monte Carlo data:
we find that theories and simulations agree in providing the picture of a fluid
exhibiting density and structural anomalies for low values of $\lambda$ and 
for both the two values of the interaction potential.
Integral equation theories give
accurate predictions for pressure and radial distribution functions, whereas
the temperatures where anomalies occur are underestimated. Upon increasing
the elongation, the RISM theory still predicts the existence of anomalies;
the latter are no longer observed in simulations, since their development
is likely precluded by the onset of crystallization. We discuss our results in
terms of the reliability of integral equation theories 
in predicting the existence of
water-like anomalies in core-softened fluids.

\end{abstract}

\maketitle

\section{Introduction}
The reference interaction site model (RISM)
theory, formulated in the early 70's by
Chandler and Andersen~\cite{chandler:72a,chandler:72b},
is still one of the most adopted structural theories of molecular
fluids. In the RISM approach, developed as a generalization of the
Ornstein-Zernike theory of atomic fluids~\cite{Hansennew}, the 
molecules are represented as a collection
of spherical interaction sites, rigidly connected 
so to reproduce a given molecular geometry~\cite{chandler:78}.
Originally the theory was developed to deal only with hard-sphere 
fluids~\cite{lowden:5228} or with simple Lennard-Jones systems~\cite{johnson}. 
Later on, the RISM framework has
been extended to take into account more realistic representations of
complex fluids, including water, methanol and other compounds 
(see Refs~\cite{lue:5427,Kovalenko:02,pettitt:7296,Kvamme:02,Munao:07} 
and a general review in Ref.~\cite{Hirata:03}).
More recently, the RISM approach (along with its
extensions and generalizations) has been 
adopted for predicting structure, thermodynamic and phase equilibria of a 
large variety of systems, including dumbbell 
fluids~\cite{Munao-cpl,Munao:PCCP,Gamez:15},
colloids~\cite{Munao:11,Tripathy-RISM}, 
proteins~\cite{Hirata:13,Hirata:15}, surfactants~\cite{Sugai:02,Kobryn:14} and
water solutions~\cite{Huang:15,Kung:10}. 
In the last years further refinements of the original RISM
theory have made possible the development of more complex approaches,
as for instance the three dimensional reference interaction site model 
theory~\cite{Miyata:08,Miyata:10,Miyata:11} and the multi-center molecular
Ornstein-Zernike equation~\cite{Sato:12}.

The structure and the phase behavior of complex
fluids is of great interest
in the field of liquid and soft matter physics.
This is expecially due to the peculiar
thermodynamic and structural properties exhibited by some of these
fluids, named anomalous
liquids. Water is the most known example of such fluids; its peculiar
behaviors, including re-entrant melting and density anomalies, have
been largely investigated both from 
experimental~\cite{debenedetti:96,soper:00,Dokter:05,clark:10,Mallamace:13} 
and theoretical~\cite{Liu:09,Poole:11,Franzese:12,Poole:13,palmer:14,smallenburg:15} points of 
view. The existence of these anomalies has been traditionally related
to the possibility to develop a network between water molecules 
through hydrogen bonds. 

Within this framework a modeling approach, which is 
arousing an increasing interest in the last years, is based on  
spherically symmetric
core-softened (CS) potentials: according to this approach, 
formulated by Hemmer and Stell in 1970~\cite{Stell:70},  
in these potentials the
hard-core interactions turn to be softened and an attractive tail is set. 
A large variety of studies has been carried out to 
investigate the peculiar physical properties of these potentials, with
a particular focus on liquid-liquid phase 
transition~\cite{Jagla:99,Malescio:01,Malescio:04,Wilding:06,Hus:14} 
and water-like anomalies~\cite{Scala:01,Franz:08,Saija:09,Frenkel:09,Presti:10,Barbosa:10,Barbosa:14,Bordin:16}. 
In this context, the RISM theory has been successfully
adopted, along with the Mode-Coupling theory, to study the pressure
dependence of the diffusion coefficient for acetonitrile and 
methanol in
water~\cite{Hirata:05} 
and to analyze the phase behavior of a number of CS molecular 
models~\cite{Urbic:14,Urbic:15}.

In this work we investigate structure and thermodynamics of a simple model
for a dimeric fluid interacting via a modified inverse-power potential 
(MIP)~\cite{Franz:JPCB,Franz:MolPhys} by means of RISM theory
and Monte Carlo (MC) simulations. In a previous 
study~\cite{Munao:16} we have shown that a system of
dimeric particles interacting through MIP
diplays typical water-like anomalies, like a temperature
of maximum density (TMD) and a non-monotonic behavior of the pair translational
entropy. Also, we have investigated how the existence 
of such anomalies depends on the elongation of the dimers and on the
interaction potential parameters. In particular, upon increasing the 
repulsion softening, it has been possible to  
study how the fluid structure is changed when going from
one-scale behavior typical of Lennard-Jones  fluids to a 
two-scale behavior characterizing the CS systems.

Here we extensively employ the RISM framework for calculating
structural and thermodynamic properties of the dimer fluid interacting
via MIP, in the whole range of elongations and for two different values 
of the repulsion softening. For this purpose we adopt and compare
hypernetted chain (HNC)~\cite{Hansennew} 
and Kovalenko-Hirata (KH)~\cite{Kovalenko:99,Kovalenko:01} approximation
closures and various routes from  the structure to thermodynamics. 
Where possible, theoretical results 
are assessed against  
previous~\cite{Munao:16} or newly generated 
MC simulations performed in the 
canonical ensemble. 
This work has been carried out with the aim to investigate the anomalous
behaviors of the dimer fluid 
in a wide range of parameters; also, to the best
of our knowledge, this study constitutes the first application of the RISM
approach to investigate density and structural anomalies in a molecular
fluid interacting via a CS potential.

The paper is organized as follows: in the next section we provide
details of the model, the RISM theory and the simulation approach. Results
are presented and discussed in the third section and conclusions follow in the
last section.

%%%%%%%%%%%%%%%%%%%%%%%%%%%%%%%%%%%%%%%%%%%%%%%%%%%%%%%%%%%%%%%%%%%%%
\begin{figure*}
\begin{center}
\begin{tabular}{ccc}
\includegraphics[width=2.4cm,angle=0]{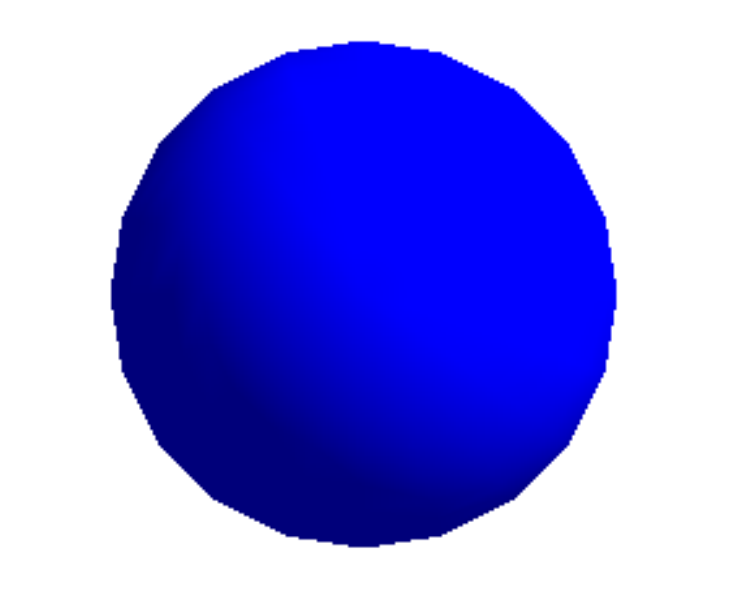} 
\includegraphics[width=3.3cm,angle=0]{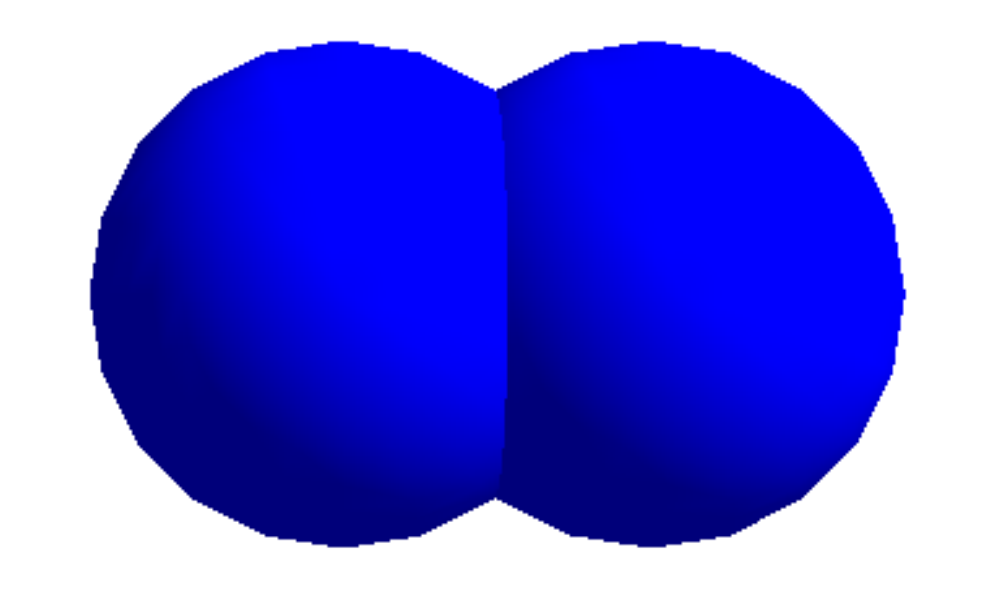} 
\includegraphics[width=3.9cm,angle=0]{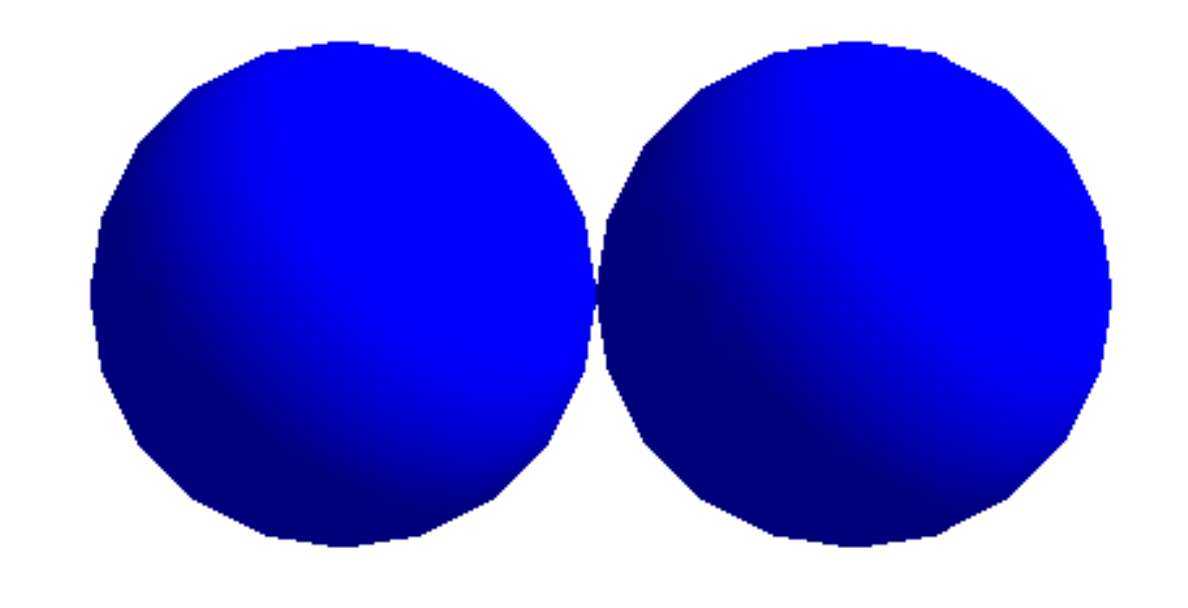} 
\end{tabular}
\caption{Cartoons of some models investigated in this work:
starting from 0 (left), $\lambda$ is progressively increased up to 
1 (right). Intermediate cases, as for instance $\lambda=0.6$ (center),
have been also investigated.}\label{fig:models}
\end{center}
\end{figure*}
%%%%%%%%%%%%%%%%%%%%%%%%%%%%%%%%%%%%%%%%%%%%%%%%%%%%%%%%%%%%%%%%%%%%%

\section{Model, theory and simulations}
A schematic representation of some models investigated in this work is 
reported
in Fig.~\ref{fig:models}: in a first instance we consider the 
monomeric case,
corresponding to two totally overlapped spheres. Then, the distance 
between the centres of the spheres is progressively increased. 
By indicating with $\lambda$ such a distance, our scheme is equivalent to
move from $\lambda=0$ (total
overlap) to $\lambda=1$ (tangent dimer). We set a
site-site interaction potential written as~\cite{Franz:JPCB}:
\begin{equation}\label{eq:pot}
U(r)=\epsilon(\sigma/r)^{n(r)} 
\end{equation}
where $r$ is the distance between spheres belonging to different dimers, 
$\epsilon$ and $\sigma$ are the units of energy and length, respectively,
and
\begin{equation}\label{eq:nr}
n(r)=n_0 \{1-\alpha\exp[-b(1-r/\sigma)^2] \} \,.
\end{equation}
In Eq.~(\ref{eq:nr}) $\alpha$ is a real number whose value is in the 
range [0, 1], whereas $b$ and $n_0$ 
are positive integer numbers. In this work we make use of the reduced units
by defining reduced temperature, density
and pressure as, respectively, $T^*\equiv k_BT/\epsilon$, 
$\rho^*\equiv \rho\sigma^3$ and $P^*\equiv P\sigma^3/\epsilon$,
where $k_{B}$ is the Boltzmann constant.
%%%%%%%%%%%%%%%%%%%%%%%%%%%%%%%%%%%%%%%%%%%%%%%%%%%%%%%%%%%%%%%%%%%%%
\begin{figure}
\begin{center}
\includegraphics[width=8.0cm,angle=0]{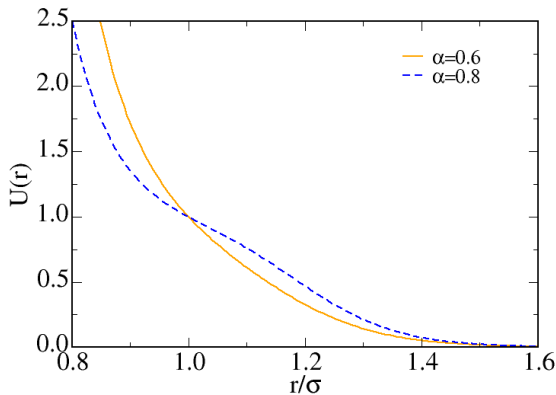} 
\caption{Site-site intermolecular potential $U(r)$ for 
$\alpha=0.6$ (full line) and for 
$\alpha=0.8$ (dashed line).}\label{fig:poten}
\end{center}
\end{figure}
%%%%%%%%%%%%%%%%%%%%%%%%%%%%%%%%%%%%%%%%%%%%%%%%%%%%%%%%%%%%%%%%%%%%%

In Eqs.~\ref{eq:pot}-\ref{eq:nr} the repulsion softening is
controlled by the parameter $\alpha$, 
whereas $b$ set the width of the interval where
$n(r)$ is smaller than $n_0$. In all calculations the values of $b=5$ and
$n_0=12$ are kept fixed; as for the values of $\alpha$, we consider two
different cases, namely $\alpha=0.6$
and $\alpha=0.8$. In Fig.~\ref{fig:poten} we report the behavior of $U(r)$ 
for these two values: as observed, for $\alpha=0.6$ the potential 
decays by following an inverse-power law, whereas for $\alpha=0.8$ an 
inflection point with a corresponding change of the concavity is found. 
Under these conditions, $U(r)$ exhibits two different length scales, as 
typical for CS systems. We remark that our interparticle potential is purely
repulsive, hence the relative
phase diagram will not exhibit a gas-liquid phase transition, even if in 
principle a metastable liquid-liquid phase transition could be 
present~\cite{Ryzhov:03}.

In order to investigate the fluid structure of the monomeric model 
(i.e. $\lambda=0$) 
we have employed the Ornstein-Zernike equation for simple 
fluids,~\cite{Hansennew} expressed in the $k$-space (for an homogeneous and
isotropic fluid) as:
%%%%%%%%%%%%%%%%%%%%%
\begin{equation}\label{eq:oz}
{\bf h}(k) = {\bf c}(k) + \rho{\bf c}(k){\bf h}(k)
\end{equation}
%%%%%%%%%%%%%%%%%%%%%
where ${\bf h}(k)$ and ${\bf c}(k)$ are the Fourier transforms 
of the total
and direct correlation functions $h(r)$ and $c(r)$
and $\rho$ is the density of the system.
In order to solve Eq.~(\ref{eq:oz}),
we have adopted in this work the
hypernetted chain (HNC) expression~\cite{Hansennew}:
%%%%%%%%%%%%
\begin{equation}\label{eq:hnc}
 c(r) = \exp[- \beta U(r) + \gamma(r)] - \gamma(r) - 1
\end{equation}
%%%%%%%%%%%%
where $\beta=1/T^*$ and $\gamma(r)=h(r) - c(r)$.
In parallel to HNC, we have employed
its partially linearized form, developed by Kovalenko and Hirata 
(KH)~\cite{Kovalenko:99,Kovalenko:01}
that amounts to set:
\begin{eqnarray}\label{eq:kh}
\qquad c(r) =
\left\{ \begin{array}{ll}
{\rm HNC} & \textrm{\qquad if \quad $g(r) \leq 1$}\\
{\rm MSA} & \textrm{\qquad if \quad $g(r) > 1$}
\end{array} \right ., 
\end{eqnarray}
%%%%%%%%%%%%
with MSA (mean spherical approximation) corresponding to 
assume~\cite{Hansennew}:
\begin{eqnarray}
\left\{ \begin{array}{ll}
{g(r)=0} & \textrm{\qquad if \quad $r \leq \sigma$}\\
{c(r)=-\beta U(r)} & \textrm{\qquad
if \quad $r > \sigma$}
\end{array} \right  .,
\end{eqnarray}
%%%%%%%%%%%%
where $g(r)=h(r) +1$.

For $\lambda > 0$,
the Ornstein-Zernike equation can not be implemented 
and a molecular generalization is required. In the present work we apply
the RISM framework~\cite{chandler:72b} where
the pair structure of a fluid composed by
identical two-site molecules
is characterized
by a set of four site-site intermolecular
pair correlation functions $h_{ij}(r)$
where $(i,j)=(1,2)$ 
The $h_{ij}(r)$
are related to a set of intermolecular direct correlation
functions $c_{ij}(r)$  by a matrix generalization of Eq.~(\ref{eq:oz}),
written in the $k$-space as:
%%%%%%%%%%%%%%%%%%%%%
\begin{equation}\label{eq:rism}
{\bf H}(k) = {\bf W}(k){\bf C}(k){\bf W}(k) +
\rho{\bf W}(k){\bf C}(k){\bf H}(k)
\end{equation}
%%%%%%%%
where
${\mathbf H}\equiv [h_{ij}(k)]$,
${\mathbf C}\equiv[c_{ij}(k)]$, and
${\mathbf W}\equiv [w_{ij}(k)]$
are $ 2 \times 2$ symmetric matrices;
the elements $w_{ij}(k)$
are the Fourier transforms
of the intramolecular correlation functions, written explicitly as:
%%%%%%%%
%\begin{eqnarray}\label{eq:intra}
\begin{equation}\label{eq:intra}
w_{ij}(k) =
\frac{\sin[kL_{ij}]}{kL_{ij}}\,,
\end{equation}
%\end{eqnarray}
%%%%%%%%
where the bond length $L_{ij}$ is given either
by  $L_{ij}=\sigma$, if $i\ne j$, or by $L_{ij}=0$, otherwise.
In analogy with the monomeric case, we have coupled Eq.~(\ref{eq:rism})
with HNC and KH approximations, generalized for molecular fluids.

In order to calculate thermodynamic properties of the monomeric fluid,
we have implemented different routes from structure to thermodynamics;
specifically, we have adopted and compared virial and compressibility 
equations of state~\cite{Hansennew,caccamo} both in HNC and KH approaches.
According to the first route, pressure can be obtained through the 
standard formula~\cite{Hansennew}:
%%%%%%%%
\begin{equation}\label{eq:vir}
\frac{\beta P}{\rho}=1-\frac{2}{3}\pi\beta\rho\int_0^\infty U'(r)g(r)r^3 dr
\end{equation}
%%%%%%%%
while in the compressibility route scheme we have~\cite{caccamo}: 
%%%%%%%%
\begin{equation}\label{eq:comp}
\beta P=\int_0^\rho d\rho' [S(k=0)]^{-1}
\end{equation}
%%%%%%%%
where $[S(k=0)]$ is the $k\to 0$ limit of the structure factor $S(k)$.
For molecular fluids, a direct application of Eq.~(\ref{eq:vir}) is not
straightforward, since the knowledge of site-site $g_{ij}(r)$ is not enough for
the calculation of the pressure. Hence, for $\lambda > 0$ we have implemented
a closed formula derived in the context of HNC 
approximation~\cite{Morita:60,Singer:85} and 
generalized for the KH closure~\cite{Kovalenko:01} as:
%%%%%%%%
\begin{eqnarray}\label{eq:pressure}
\frac{\beta P}{\rho} & = & 1 +\frac{\rho}{2} \sum_{ij}
\int \diff{\bf r} \left[
\frac{1}{2} h_{ij}^{2}(r) \Theta(-h_{ij}(r))
- c_{ij}(r) \right]   \nonumber\\[4pt]
& & \ \  +
\frac{1}{2(2\pi)^{3}} \int \diff{\bf k}
\{\rho^{-1} \ln \det\,[{\mathbf I} -
\rho {\mathbf W}(k) {\mathbf C}(k)] \nonumber\\[4pt]
& & \ \ \ \   - {\rm Tr}\,
[{\mathbf W}(k) {\mathbf C}(k)] [{\mathbf I} -
\rho {\mathbf W}(k) {\mathbf C}(k)]^{-1}\}\,
\end{eqnarray}
%%%%%%%%
where $\Theta$ is the Heaviside step function.
As for the compressibility route (see Eq.~(\ref{eq:comp})) 
its validity is guaranteed
in a molecular context also.
%%%%%%%%%%%%%%%%%%%%%%%%%%%%%%%%%%%%%%%%%%%%%%%%%%%%%%%%%%%%%%%%%%%%%
\begin{figure}
\begin{center}
\includegraphics[width=8.0cm,angle=0]{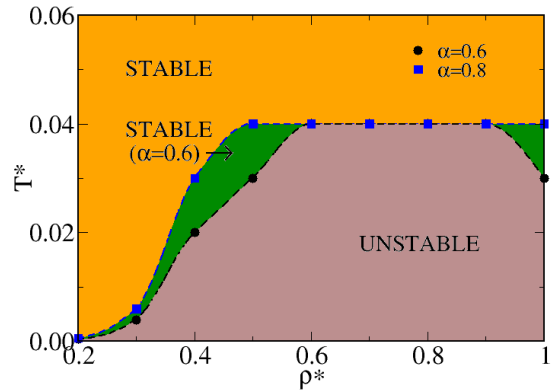} 
\caption{Stability regions of the HNC algorithm for $\lambda=0$.
In the orange area the numerical convergence is achieved both for $\alpha=0.6$
and for 
$\alpha=0.8$; in the green zone, the convergence is ensured for 
$\alpha=0.6$ only, whereas in the brown area the HNC fails to get the 
convergence regardless of the value of $\alpha$.}\label{fig:hnc}
\end{center}
\end{figure}
%%%%%%%%%%%%%%%%%%%%%%%%%%%%%%%%%%%%%%%%%%%%%%%%%%%%%%%%%%%%%%%%%%%%%

We have implemented the numerical solution of both Ornstein-Zernike and RISM
schemes by means of  a standard iterative Picard algorithm,
on a grid of 8192 points with a mesh $\Delta r=0.005\sigma$.
To improve the convergence of the Picard algorithm we have 
adopted the strategy of mixing old and new $\gamma(r)$ functions with a 
mixing parameter of 0.9. This choice allows for an optimization of the
convergence procedure for a wide range of temperatures and densities.

Theoretical predictions have been compared with Monte Carlo
(MC) simulations performed in the canonical ensemble. We have considered 
a system composed by 864 dimers in a cubic box with periodic boundary 
conditions. 
For any state point we have first performed $2 \times 10^5$
steps in order to equilibrate the system, then computing statistical
averages on the same number of following steps. For the lowest 
temperatures investigated, we have employed up to $5 \times 10^5$ steps for the 
equilibration stage; then, an equal number of steps has been generated in the
production stage.
%%%%%%%%%%%%%%%%%%%%%%%%%%%%%%%%%%%%%%%%%%%%%%%%%%%%%%%%%%%%%%%%%%%%%
\begin{figure}
\begin{center}
\includegraphics[width=8.0cm,angle=0]{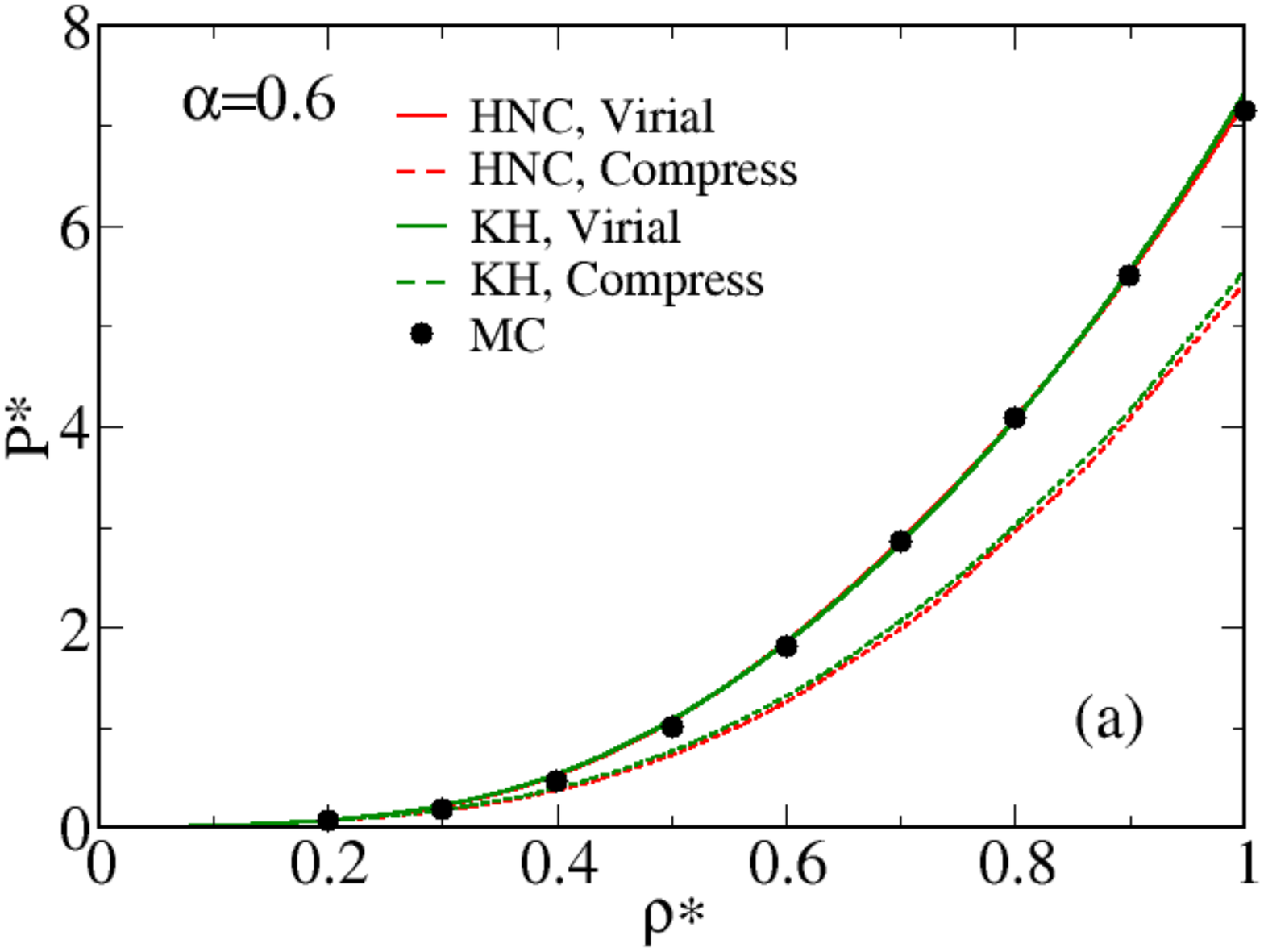} \\
\includegraphics[width=8.0cm,angle=0]{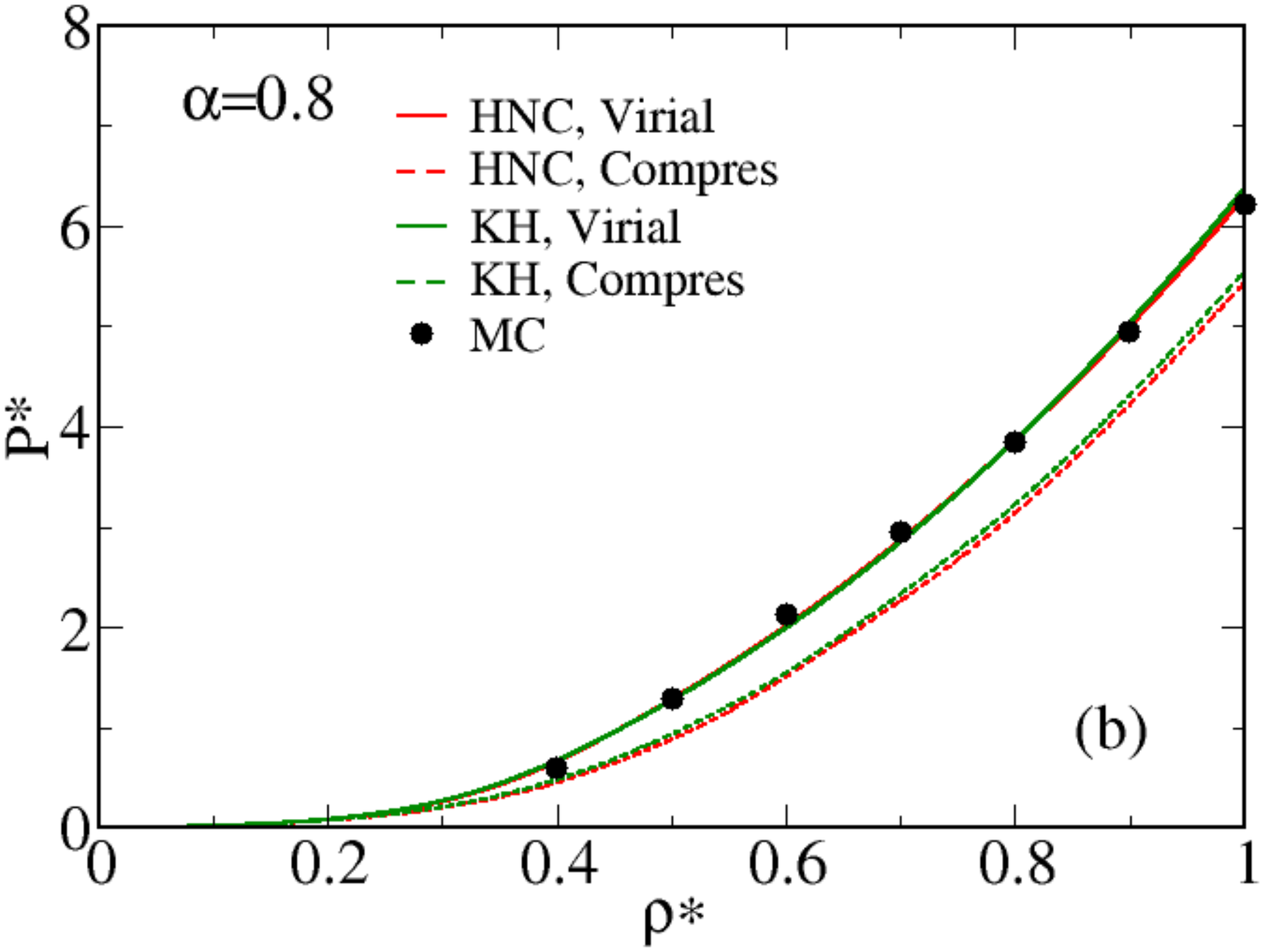} 
\caption{HNC (red lines), KH (green lines) and MC (symbols) pressures 
for $\lambda=0$, $T^*=0.10$
and $\alpha=0.6$ (a) and $\alpha=0.8$ (b). Virial and compressibility routes
are indicated by full lines and dashed lines, respectively.}\label{fig:EOS}
\end{center}
\end{figure}
%%%%%%%%%%%%%%%%%%%%%%%%%%%%%%%%%%%%%%%%%%%%%%%%%%%%%%%%%%%%%%%%%%%%%

\section{Results}
%%%%%%%%%%%%%%%%%%%%%%%%%%%%%%%%%%%%%%%%%%%%%%%%%%%%%%%%%%%%%%%%%%%%%
We have first considered the case $\lambda=0$. As a preliminary check,
we have identified the regions, in the temperature-density plane, where
the HNC numerical algorithm is able to achieve the convergence. 
Such regions are
reported in Fig.~\ref{fig:hnc}: the algorithm properly works at
all densities for $T^*\geq 0.04$. At lower temperatures, the numerical 
procedure encounters more and more difficulties to get the fully convergence,
until it ceases to work; for $\alpha=0.8$ this breakdown is observed at 
temperatures slightly higher than for $\alpha=0.6$. 
%Such a circumstance
%suggests that the increase of $\alpha$ causes a slight anticipation 
%of the destabilization of
%the numerical procedure when very low temperatures are approached.
Conversely, we have verified that the KH closure does not show any
convergence problems for $T^*\geq 0.02$ in the whole range of densities and
for both $\alpha=0.6$ and $\alpha=0.8$.  This circumstance allows for the
implementation of the KH closure even at low temperatures where HNC is not
able to provide predictions.

%%%%%%%%%%%%%%%%%%%%%%%%%%%%%%%%%%%%%%%%%%%%%%%%%%%%%%%%%%%%%%%%%%%%%
\begin{figure}
\begin{center}
\includegraphics[width=8.0cm,angle=0]{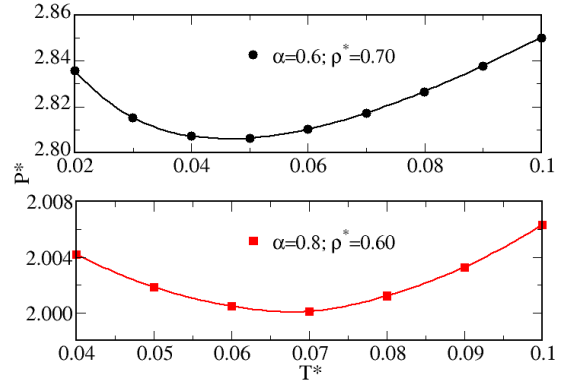} \\
\caption{Pressures versus temperatures for the monomeric case for
$\alpha=0.6$ and $\rho^*=0.70$ (top) and for 
$\alpha=0.8$ and $\rho^*=0.60$ (bottom), resulting  
from the HNC virial route.}\label{fig:PvsT}
\end{center}
\end{figure}
%%%%%%%%%%%%%%%%%%%%%%%%%%%%%%%%%%%%%%%%%%%%%%%%%%%%%%%%%%%%%%%%%%%%%

Simulation and theoretical pressures for $\alpha=0.6$ and 
$\alpha=0.8$ at $T^*=0.10$ are reported in panels (a) and (b) of 
Fig.~\ref{fig:EOS}, respectively. Ornstein-Zernike predictions 
have been obtained by following the
virial equation (see Eq.~(\ref{eq:vir})) and the compressibility 
route (see Eq.~(\ref{eq:comp})). Both HNC and KH show a quantitative
agreement with the MC data if the virial route is implemented; conversely,
the compressibility route systematically underestimates simulation 
results in 
the whole density range. Moreover, at this temperature 
no remarkable differences between HNC and KH are observed.  
We have also verified that simulation data for internal energy per 
particle $E/N$, not shown here, are accurately reproduced by both theories, 
with only a slight overestimation at intermediate densities. 

Upon lowering $T^*$ 
we document an unusual expansion of the system, with the density progressively
increasing until it reaches a maximum and then decreasing. Such a scenario
points out to the presence of a temperature of maximum density (TMD), whose 
value strongly depends on the particular pressure considered.  
As a consequence, if we look at the behavior of the pressure
as a function of the temperature, a minimum is observed at a specific value
of the density. Such a minimum is reported in Fig.~\ref{fig:PvsT} in the 
context of HNC virial like for both $\alpha=0.6$ and $\alpha=0.8$ at 
different densities: we observe
that for $\alpha=0.6$ the minimum is located at 
lower temperatures and higher pressures. 
This is consistent with the fact that for $\alpha=0.6$ 
the TMD is very close to the liquid-solid transition~\cite{Franz:MolPhys}. 

%%%%%%%%%%%%%%%%%%%%%%%%%%%%%%%%%%%%%%%%%%%%%%%%%%%%%%%%%%%%%%%%%%%%%
\begin{figure}
\begin{center}
\includegraphics[width=8.0cm,angle=0]{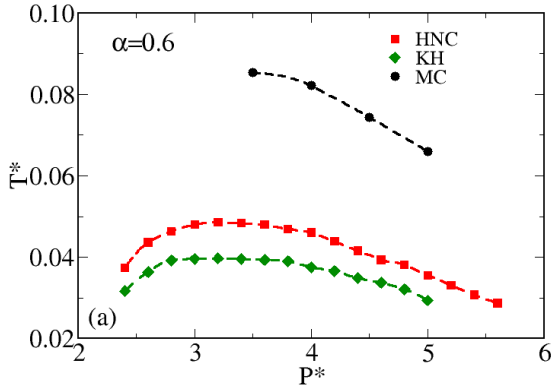} \\
\includegraphics[width=8.0cm,angle=0]{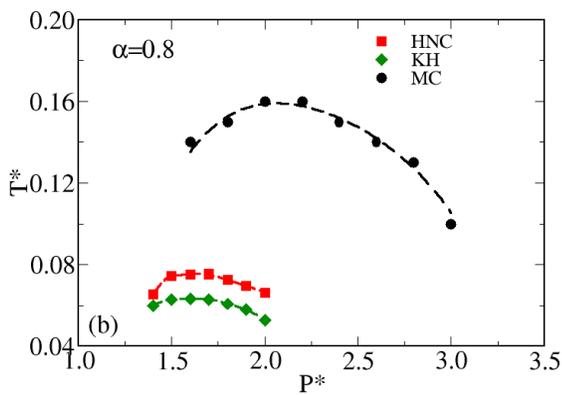} 
\caption{Loci of TMD points for $\lambda=0$ with 
$\alpha=0.6$ (a) and 
$\alpha=0.8$ (b) resulting  
from HNC, KH and MC. Theoretical predictions have been obtained by the
virial route. Lines are guides for the eye.}\label{fig:route}
\end{center}
\end{figure}
%%%%%%%%%%%%%%%%%%%%%%%%%%%%%%%%%%%%%%%%%%%%%%%%%%%%%%%%%%%%%%%%%%%%%

The specific values attained by the TMD define the density anomaly region 
in the temperature-pressure plane. Such regions are reported in 
Fig.~\ref{fig:route}, where theoretical results have been
obtained by employing the virial route only; 
we have checked that at low temperatures and high densities 
the HNC compressibility route can hardly be implemented, since the integration
of $[S(k=0)]^{-1}$ shows typical oscillations~\cite{Urbic:14} that affect
the behavior of the resulting TMD curves, not shown here. 
In Fig.~\ref{fig:route} we note that, despite the quantitative agreement between
pressures obtained from MC simulations and from the 
virial route 
shown in Fig.~\ref{fig:EOS}, the loci of TMD points are 
underestimated by both KH and HNC, with the latter slightly more predictive.
This outcome can be explained by considering that, since the TMD is strongly
dependent on the pressure, a tiny difference between the pressure values 
may give rise to a remarkable difference between the TMD's. 
It is worth to compare our results 
with previous integral equations studies
of anomalies in simple fluids~\cite{Oliveira:06,Lomba:07}. In such
studies the anomalies have not been observed in the HNC framework, but only
in the more refined context of thermodynamically 
consistent closures, like the
Rogers-Young. However, the intermolecular potentials adopted in those studies
included also attractive contributions and it is known that the HNC closure 
hardly deals with attractive potentials in proximity of phase transitions. Hence
we desume that HNC is able to detect the presence of fluid anomalies of the 
MIP potential because the latter does not contain attractive contributions and
the anomalous region is located inside the convergence region of the 
algorithm.

%%%%%%%%%%%%%%%%%%%%%%%%%%%%%%%%%%%%%%%%%%%%%%%%%%%%%%%%%%%%%%%%%%%%%
\begin{figure}
\begin{center}
\includegraphics[width=8.0cm,angle=0]{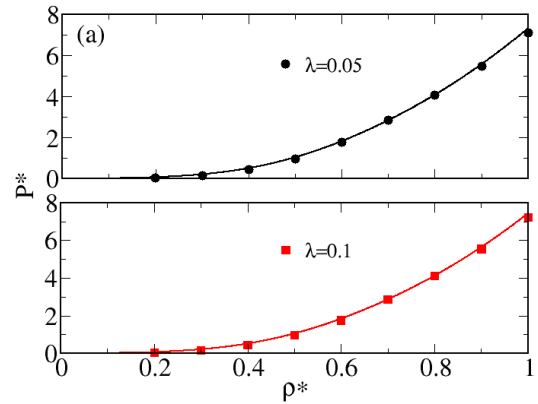} \\
\includegraphics[width=8.0cm,angle=0]{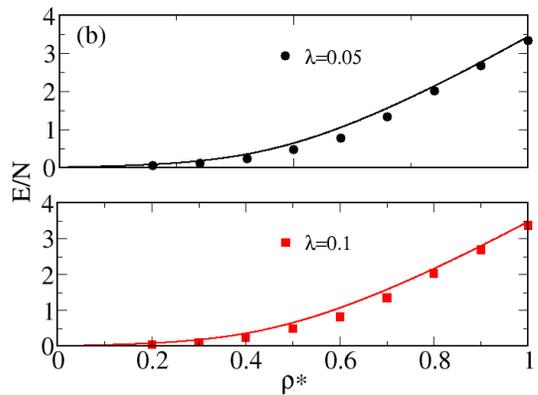} 
\caption{Pressure (a) and 
average internal energy per particle (b) for $\alpha=0.6$,
$T^*=0.075$ and $\lambda=0.05$ and 0.1, obtained from KH (lines)
and MC (symbols).}\label{fig:a06}
\end{center}
\end{figure}
%%%%%%%%%%%%%%%%%%%%%%%%%%%%%%%%%%%%%%%%%%%%%%%%%%%%%%%%%%%%%%%%%%%%%
Upon increasing $\lambda$, the Ornstein-Zernike approach must be replaced by
the RISM framework. We have verified that the HNC scheme fails to achieve the
convergence at low temperatures for $\lambda < 0.2$; such a circumstance
precludes the possibility to implement this closure to investigate structural
and thermodynamic anomalies for 
low values of $\lambda$. Henceforth we shall
make use of the KH closure only, that does not experience such problems
and properly works in the whole range of $\lambda$ values. Also,
following the prescription adopted in Ref.~\cite{Munao:16}
and previously suggested by de Oliveira and coworkers~\cite{Barbosa:10}, 
pressure and temperature shall be rescaled by a factor 4, in order to 
ensure a proper comparison with the monomeric case where the effective
interparticle interaction is four times weaker.

In Fig.~\ref{fig:a06} we compare KH and MC results for 
pressure (a) and internal energy per particle (b) 
for $\alpha=0.6$, $T^*=0.075$
and $\lambda=0.05$ and 0.1. Theoretical predictions for the pressure 
have been obtained
by implementing the KH closed formula~\cite{Kovalenko:01} 
reported in Eq.~(\ref{eq:pressure}).
%%%%%%%%%%%%%%%%%%%%%%%%%%%%%%%%%%%%%%%%%%%%%%%%%%%%%%%%%%%%%%%%%%%%%
\begin{figure}
\begin{center}
\includegraphics[width=8.0cm,angle=0]{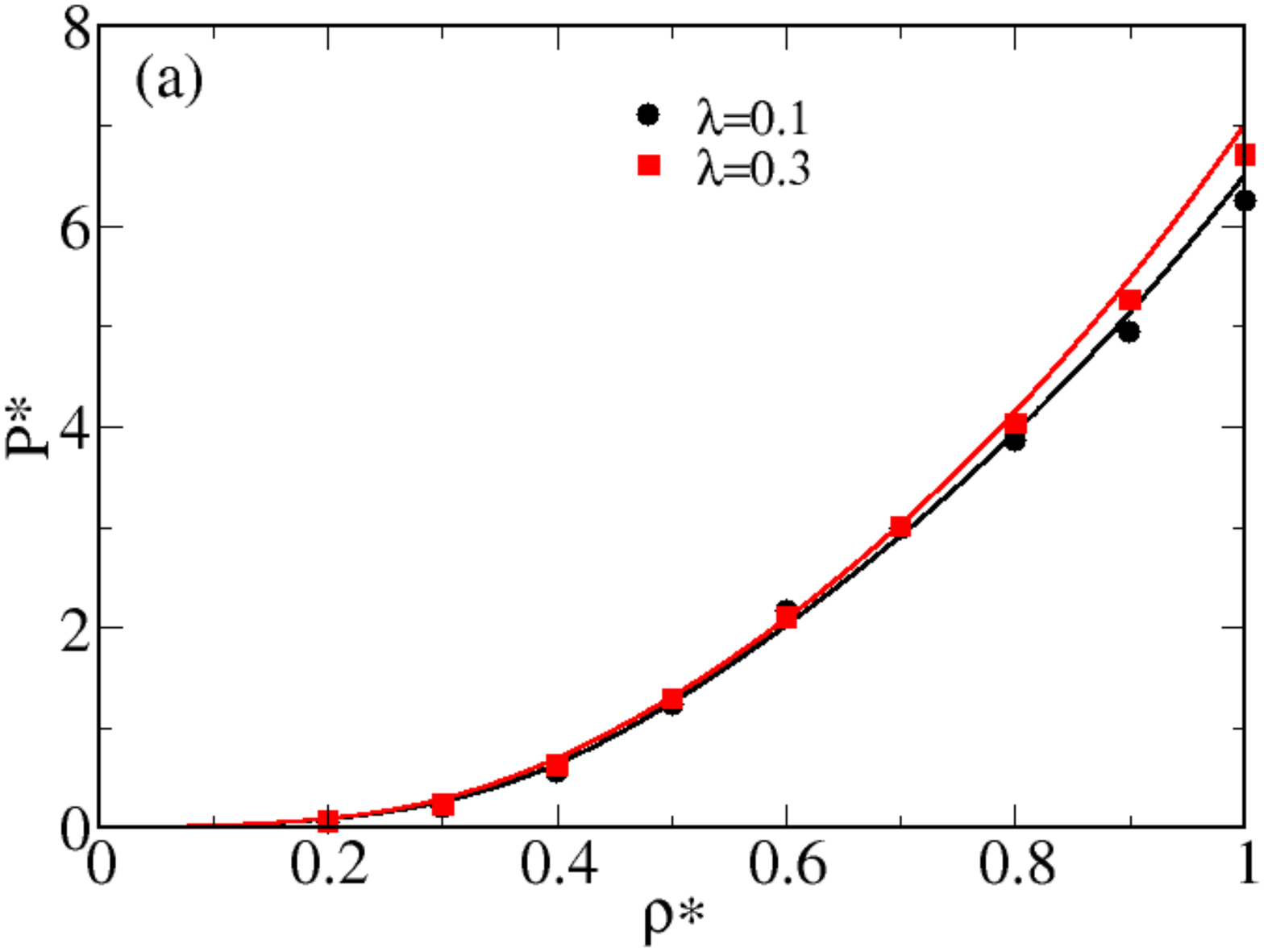} \\
\includegraphics[width=8.0cm,angle=0]{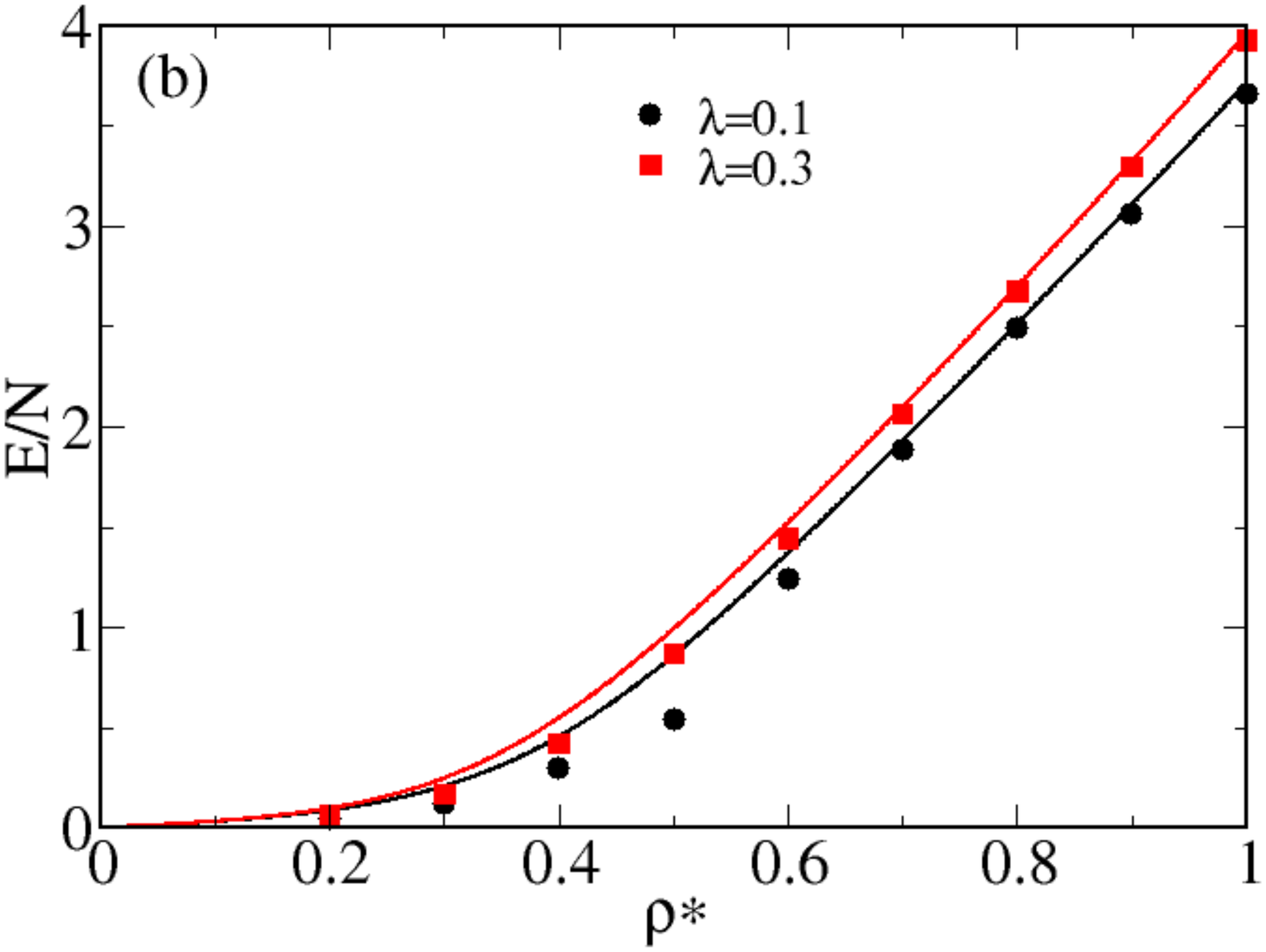} 
\caption{Pressure (a) and 
average internal energy per particle (b) for $\alpha=0.8$, 
$T^*=0.075$ and $\lambda=0.1$ and 0.3, obtained from KH (lines)
and MC (symbols).}\label{fig:a08}
\end{center}
\end{figure}
%%%%%%%%%%%%%%%%%%%%%%%%%%%%%%%%%%%%%%%%%%%%%%%%%%%%%%%%%%%%%%%%%%%%%
%%%%%%%%%%%%%%%%%%%%%%%%%%%%%%%%%%%%%%%%%%%%%%%%%%%%%%%%%%%%%%%%%%%%%
\begin{figure}
\begin{center}
\includegraphics[width=8.0cm,angle=0]{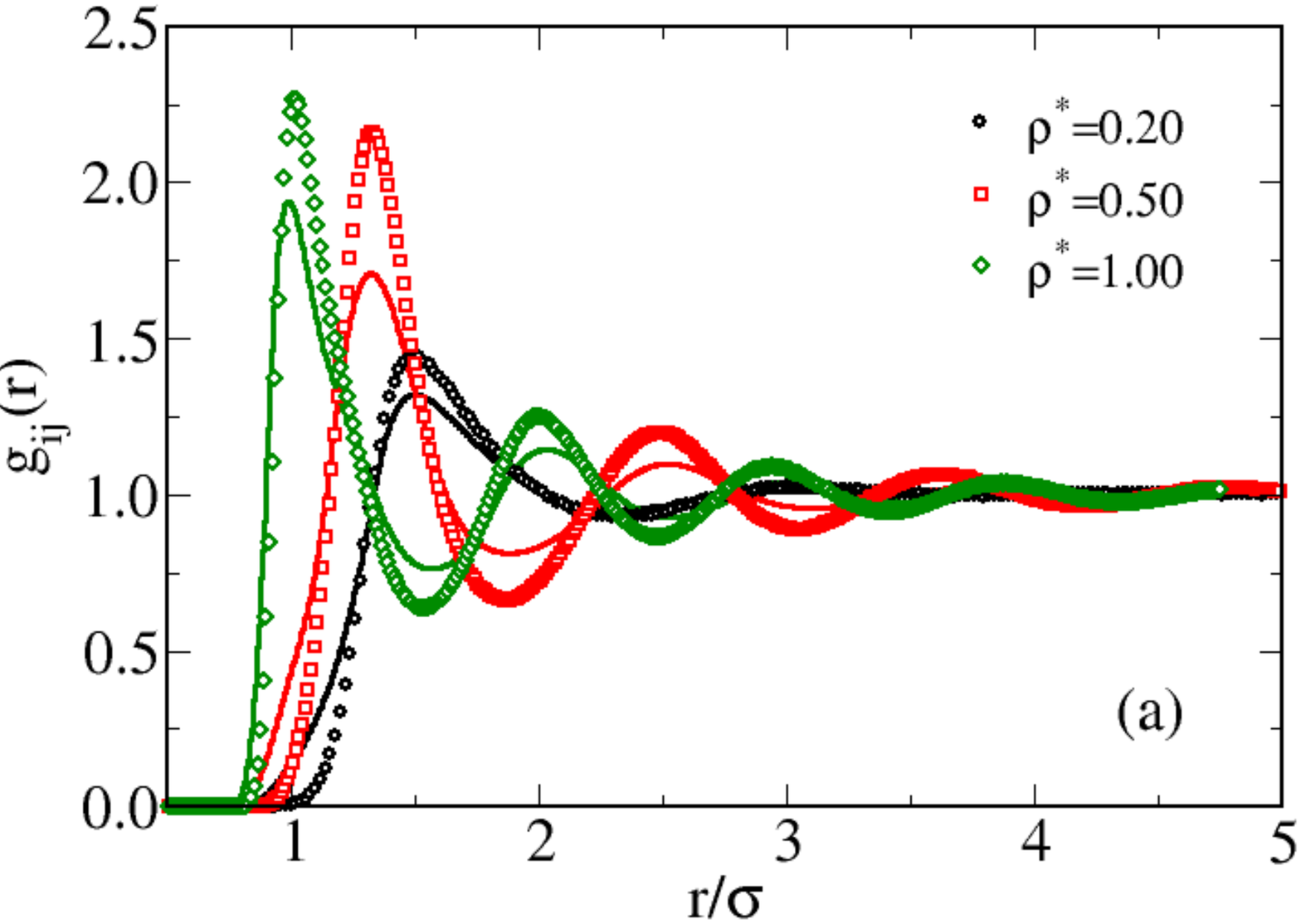} \\
\includegraphics[width=8.0cm,angle=0]{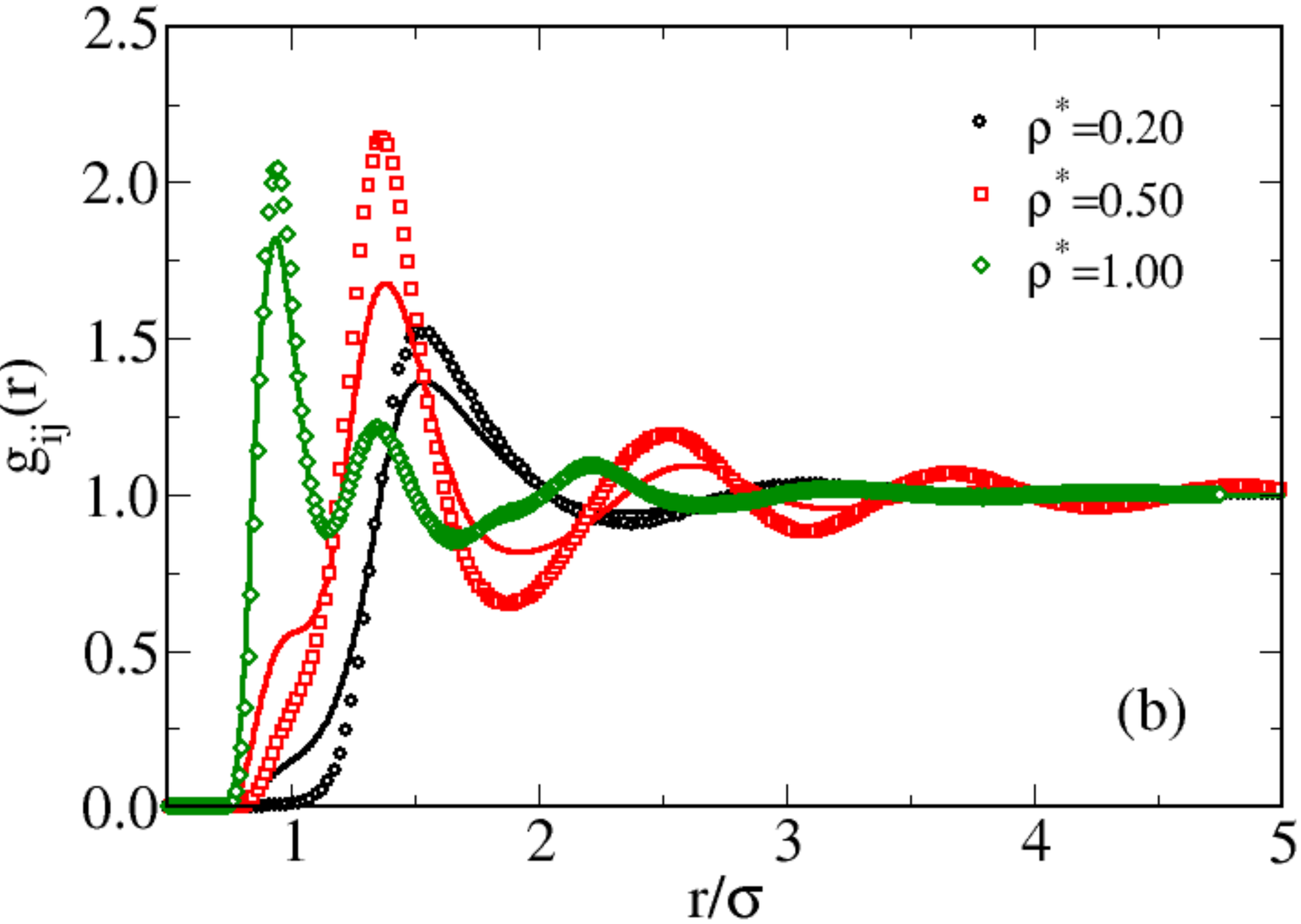} 
\caption{Site-site $g_{ij}(r)$ for $\alpha=0.6$ and 
$\lambda=0.05$ (a) and for $\alpha=0.8$ and $\lambda=0.1$ (b)
at $T^*=0.15$ and various densities
obtained from KH (lines) and MC (symbols).}\label{fig:gr}
\end{center}
\end{figure}
%%%%%%%%%%%%%%%%%%%%%%%%%%%%%%%%%%%%%%%%%%%%%%%%%%%%%%%%%%%%%%%%%%%%%
As already observed for the
monomeric case, pressure is still quantitatively reproduced by the theory
both for $\lambda=0.05$ and for $\lambda=0.1$. Theoretical predictions for 
the
internal energy appear less accurate, with an overestimation visible at
intermediate densities. Upon increasing $\alpha$ to 0.8, we find a similar 
scenario: in Fig.~\ref{fig:a08} we report simulation and theoretical results
for pressure (a) and internal energy per particle (b)
for $\alpha=0.8$, $T^*=0.075$ and $\lambda=0.1$ and 0.3. 
The theory accurately reproduces simulation results for pressure,
but for high
densities ($\rho^* > 0.8$) where KH slightly overestimates MC data.
The agreement worsens for the internal energy, in particular at intermediate
densities as observed for $\alpha=0.6$.

The local structure of the dimeric fluid
is investigated in Fig.~\ref{fig:gr} where we report
KH and MC results for the site-site radial distribution function $g_{ij}(r)$
for $\alpha=0.6$ and $\lambda=0.05$ (a) and for $\alpha=0.8$ and 
$\lambda=0.1$ (b) at $T^*=0.15$ and various densities. The agreement 
between theory and simulations appears reasonably good in both cases: for
$\alpha=0.6$ the one-scale behavior of the intermolecular potential 
(see Fig.~\ref{fig:poten}) is clearly visible in the $g_{ij}(r)$. Indeed,
upon increasing the density, the first peak is progressively shifted towards
lower and lower values of $r/\sigma$, as expected for a core-softened
fluid subjected to a compression. KH closely follows the simulation path,
becoming more predictive at high density ($\rho^*=1.0$). For $\alpha=0.8$
(b) the behavior of $g_{ij}(r)$ appears quite different; now, a second
close-contact peak is found upon increasing the density. Such a peak 
becomes dominant at $\rho^*=1.0$ where the system is strongly compressed.
This behavior is typical of a fluid exhibiting two length scales in the 
interaction potential and is observed both in KH and MC $g_{ij}(r)$, with 
the theory anticipating the onset of the second peak, already visible 
for $\rho^*=0.50$.

%%%%%%%%%%%%%%%%%%%%%%%%%%%%%%%%%%%%%%%%%%%%%%%%%%%%%%%%%%%%%%%%%%%%%
\begin{figure}
\begin{center}
\includegraphics[width=8.0cm,angle=0]{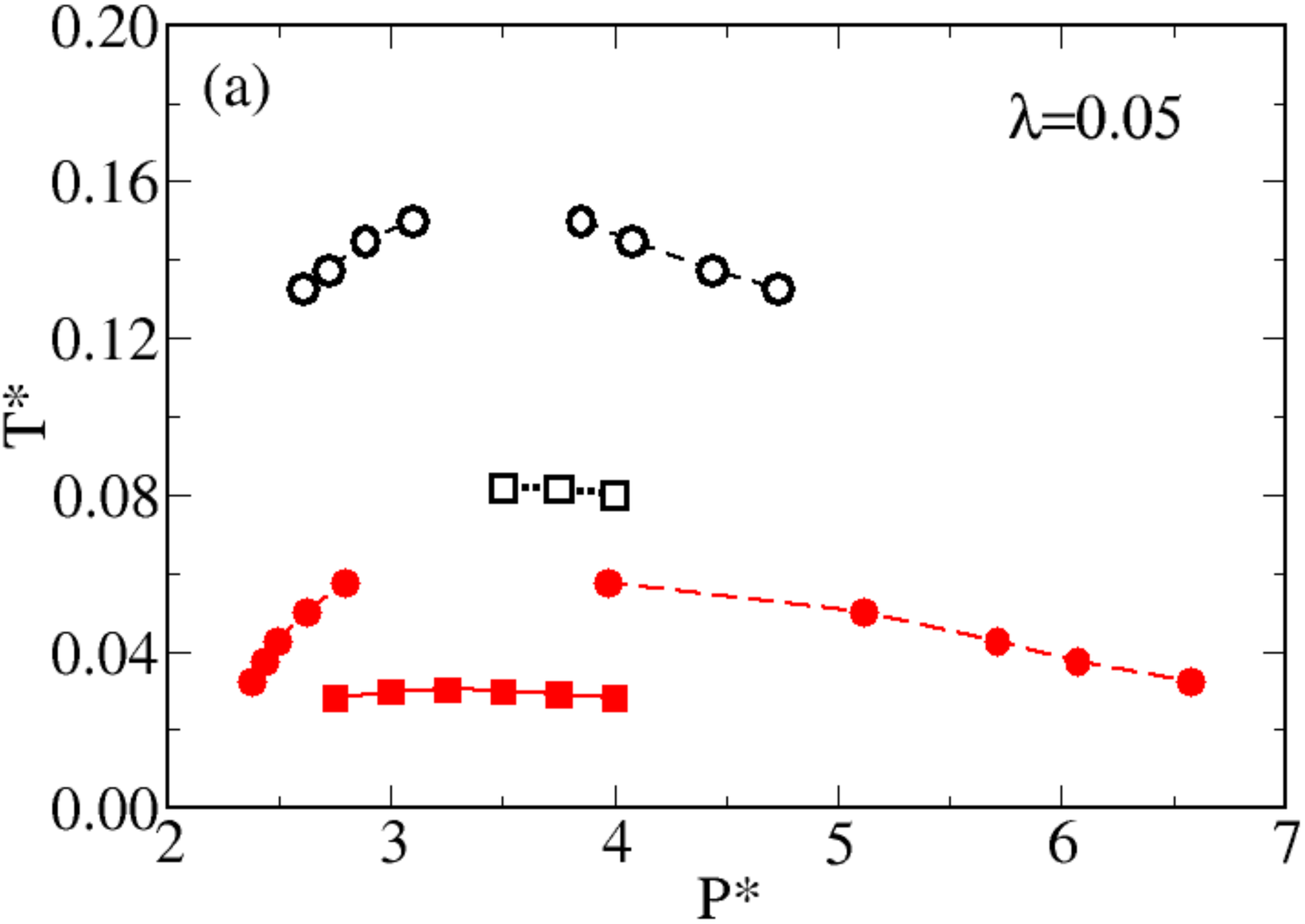} \\
\includegraphics[width=8.0cm,angle=0]{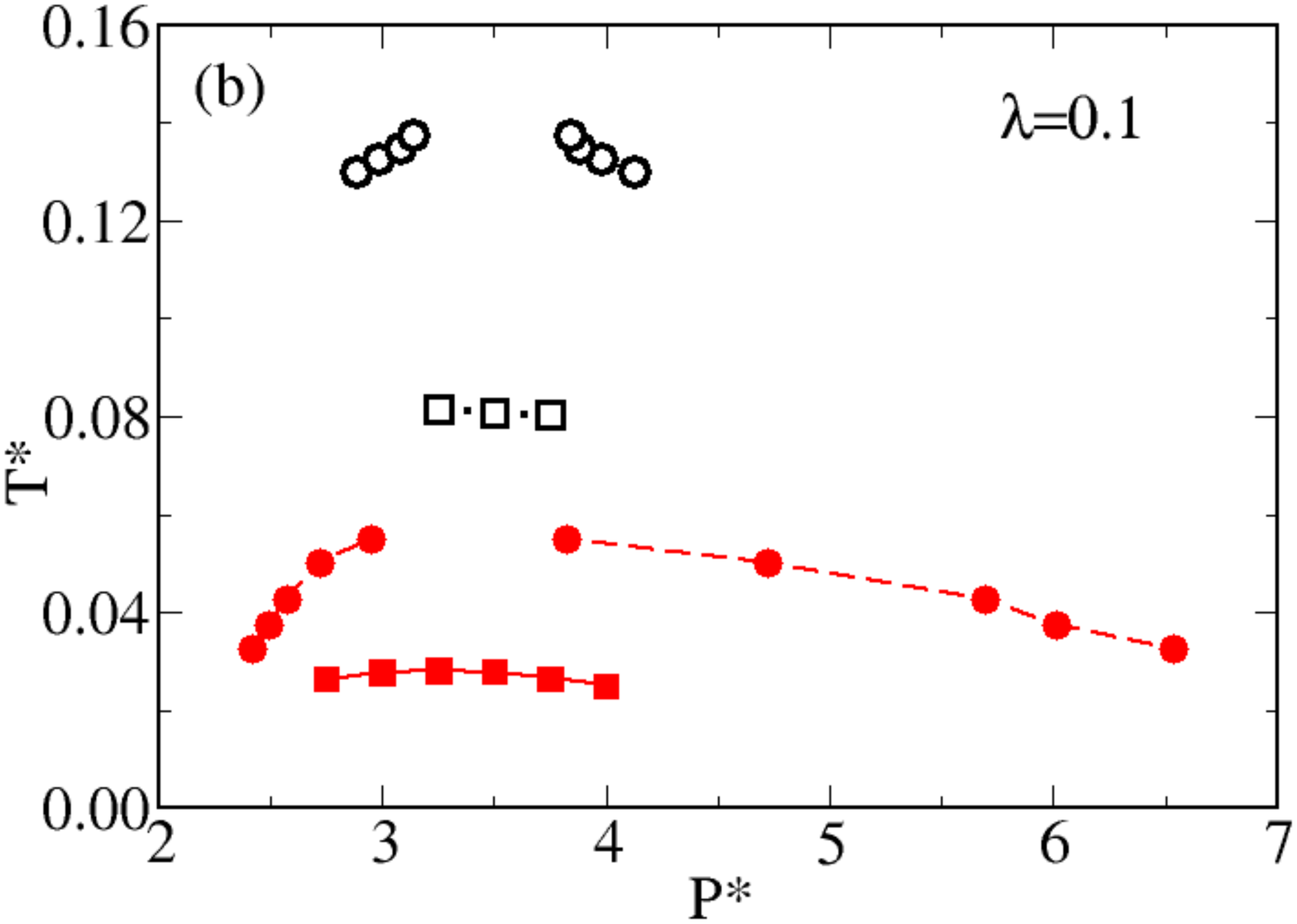} \\
\caption{Loci of structural (circles) and density (squares) anomalies 
in the pressure-temperature plane for $\alpha=0.6$ and 
increasing $\lambda$
obtained from KH (full symbols) and MC (open symbols).
Lines are guides for the eye.}\label{fig:anom-a06}
\end{center}
\end{figure}
%%%%%%%%%%%%%%%%%%%%%%%%%%%%%%%%%%%%%%%%%%%%%%%%%%%%%%%%%%%%%%%%%%%%%
%%%%%%%%%%%%%%%%%%%%%%%%%%%%%%%%%%%%%%%%%%%%%%%%%%%%%%%%%%%%%%%%%%%%%
\begin{figure}
\begin{center}
\includegraphics[width=7.8cm,angle=0]{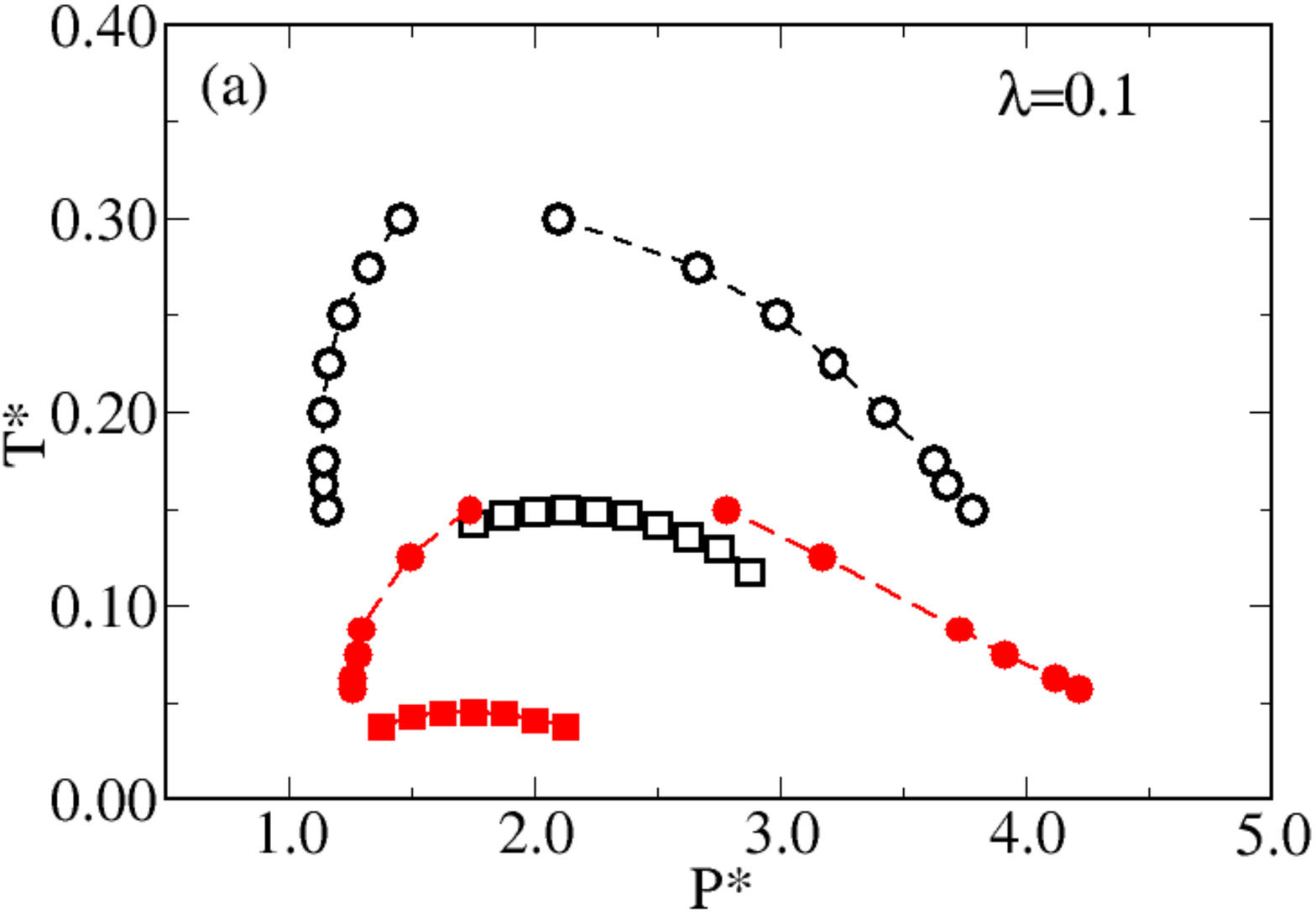} \\
\includegraphics[width=7.8cm,angle=0]{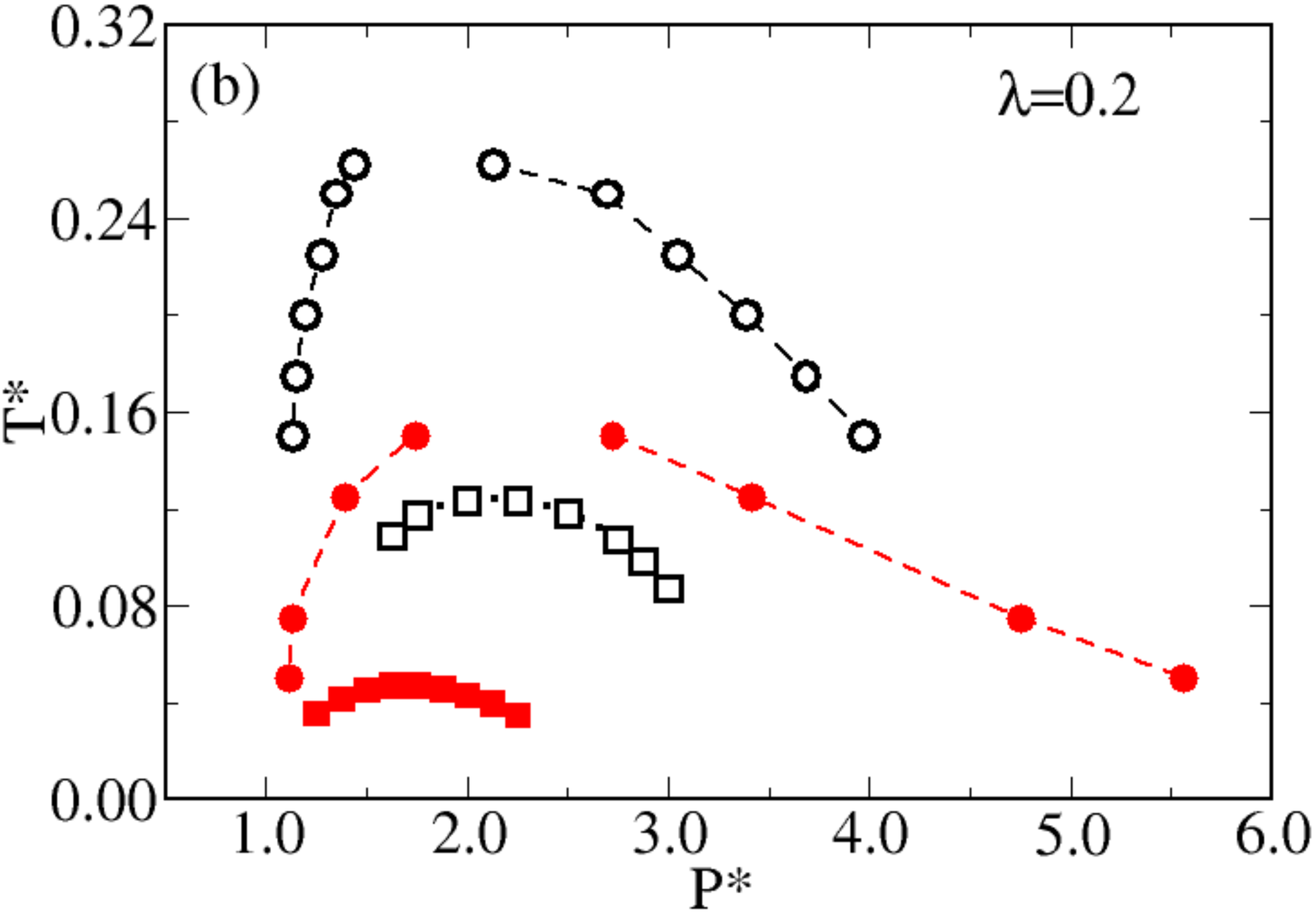} \\
\includegraphics[width=7.8cm,angle=0]{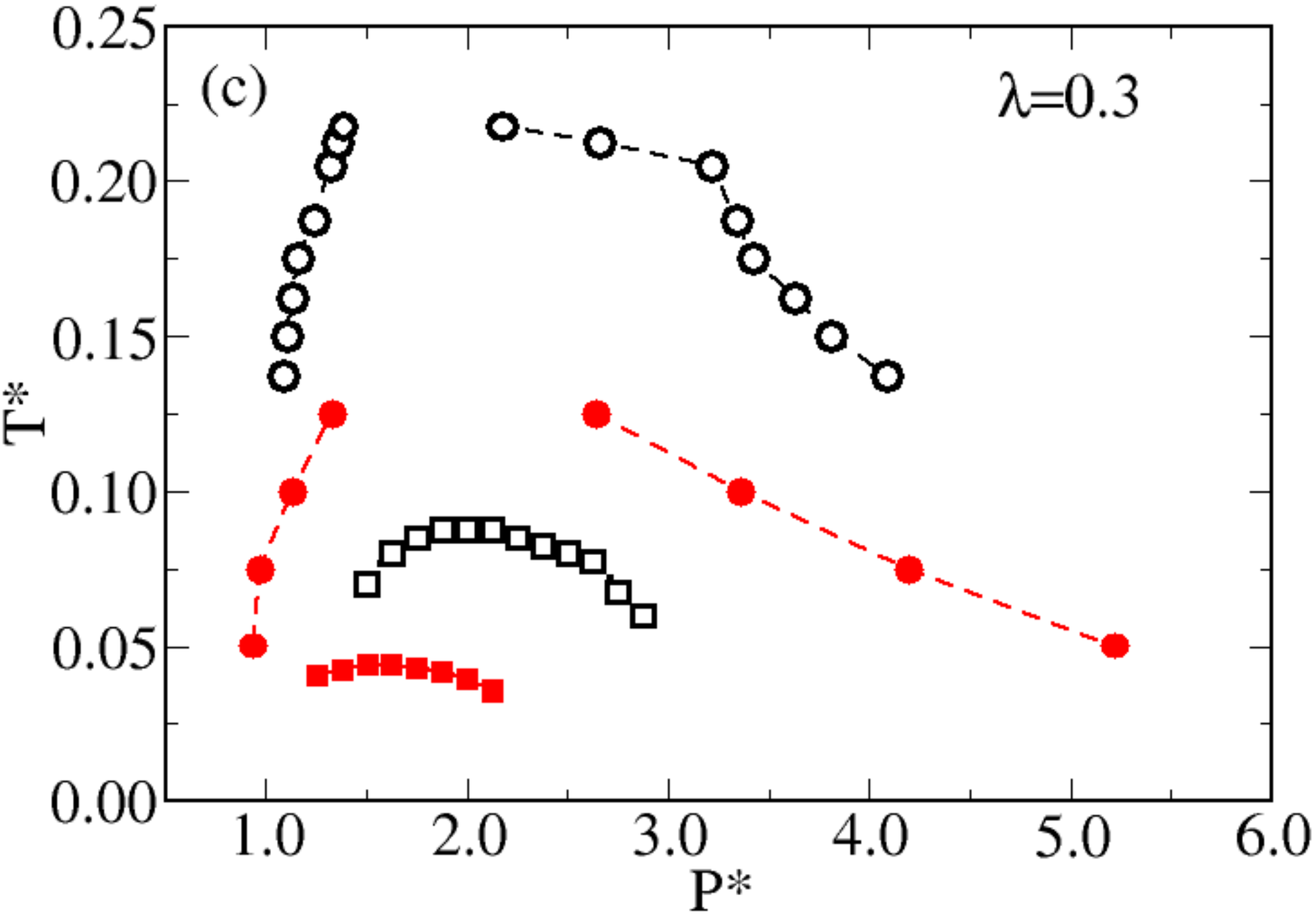} 
\caption{Loci of structural (circles) and density (squares) anomalies 
in the pressure-temperature plane for $\alpha=0.8$ and 
increasing $\lambda$
obtained from KH (full symbols) and MC (open symbols).
Lines are guides for the eye.}\label{fig:anom-a08}
\end{center}
\end{figure}
%%%%%%%%%%%%%%%%%%%%%%%%%%%%%%%%%%%%%%%%%%%%%%%%%%%%%%%%%%%%%%%%%%%%%

We now tackle the issue to investigate the behavior of density and 
structural anomalies as a function of $\lambda$ for $\alpha=0.6$
and for $\alpha=0.8$. As previously discussed, the density
anomaly region in the temperature-pressure plane is defined by the values
attained by the TMD. The structural anomaly is instead characterized by
the unusual behavior of the pair translational entropy $S_2$, defined 
as~\cite{Green}:
%%%%%%%%%%%%%%%%%%%%%%%%%%%%%%%%%%%%%%%%%%%%%%%%%%%%%%%%%%%%%%%%%%%%%
\begin{equation}\label{eq:s2}
S_2/k_{B}=-\frac{1}{2}\rho\int d{\bf r} [g_{cm}(r) {\rm ln}  g_{cm}(r) -g_{cm}(r) +1]
\end{equation}
%%%%%%%%%%%%%%%%%%%%%%%%%%%%%%%%%%%%%%%%%%%%%%%%%%%%%%%%%%%%%%%%%%%%%
where $g_{cm}(r)$ is the pair distribution function between the centers of
mass of two dimers. 
This quantity is not to be taken as a 
quantitative measure of the pair entropy of the dimer because the 
rotational contribution is not considered by Eq.~(\ref{eq:s2}). 
In fact, the orientational term depends on the angles 
that are needed to specify 
the relative orientation. The correlation between the 
translational and orientational contributions of the pair entropy turns 
out to be a sensitive indicator of structurally resolved ordering process 
occurring in molecular liquids~\cite{karplus:96,saitta:03,esposito:06,
huggins:12,kuffel:14} 
and it will be subject of a subsequent study~\cite{munao:17}. 
%%%%%%%%%%%%%%%%%%%%%%%%%%%%%%%%%%%%%%%%%%%%%%%%%%%%%%%%%%%%%%%%%%%%%
\begin{figure}
\begin{center}
\includegraphics[width=8.0cm,angle=0]{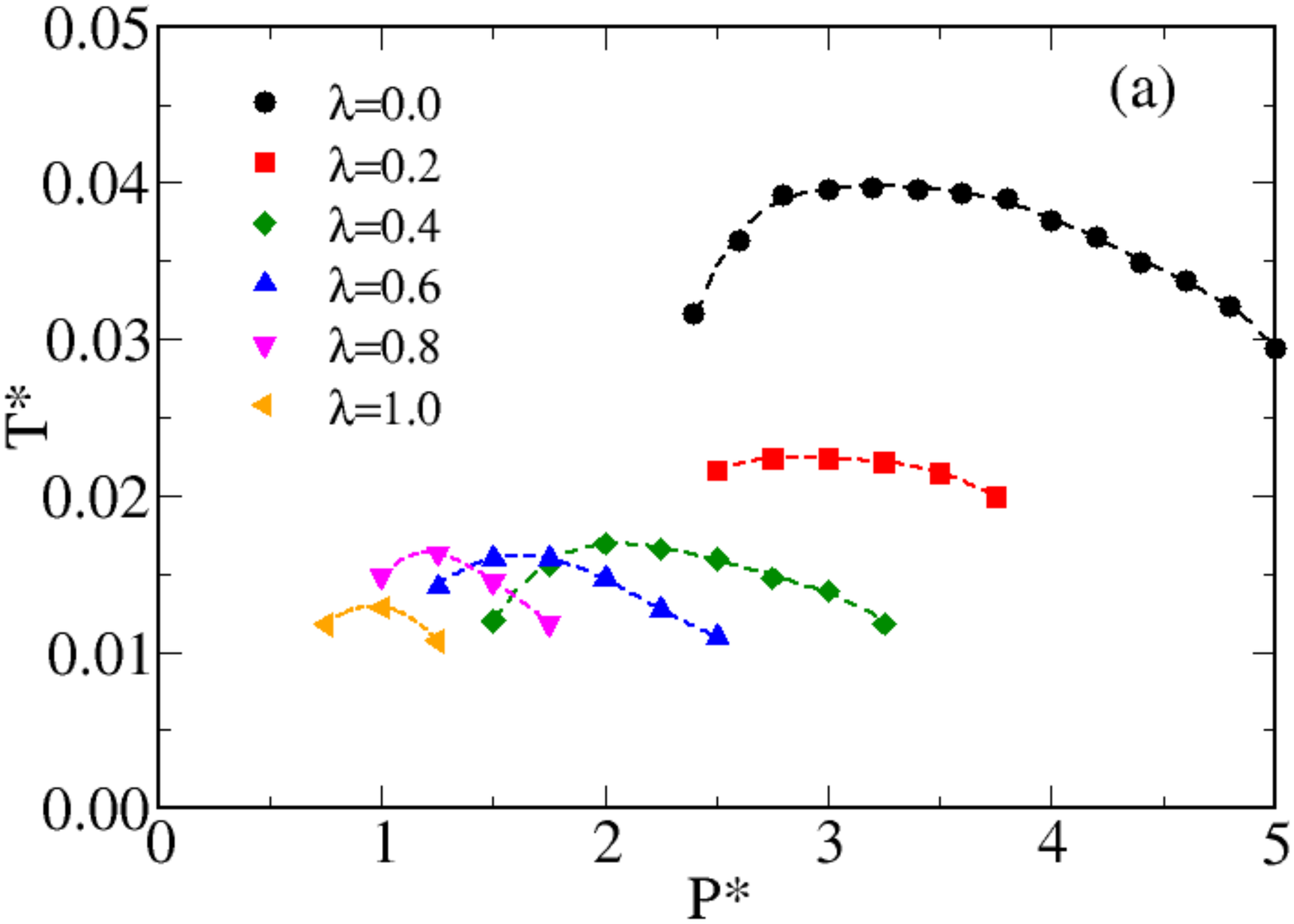} \\
\includegraphics[width=8.0cm,angle=0]{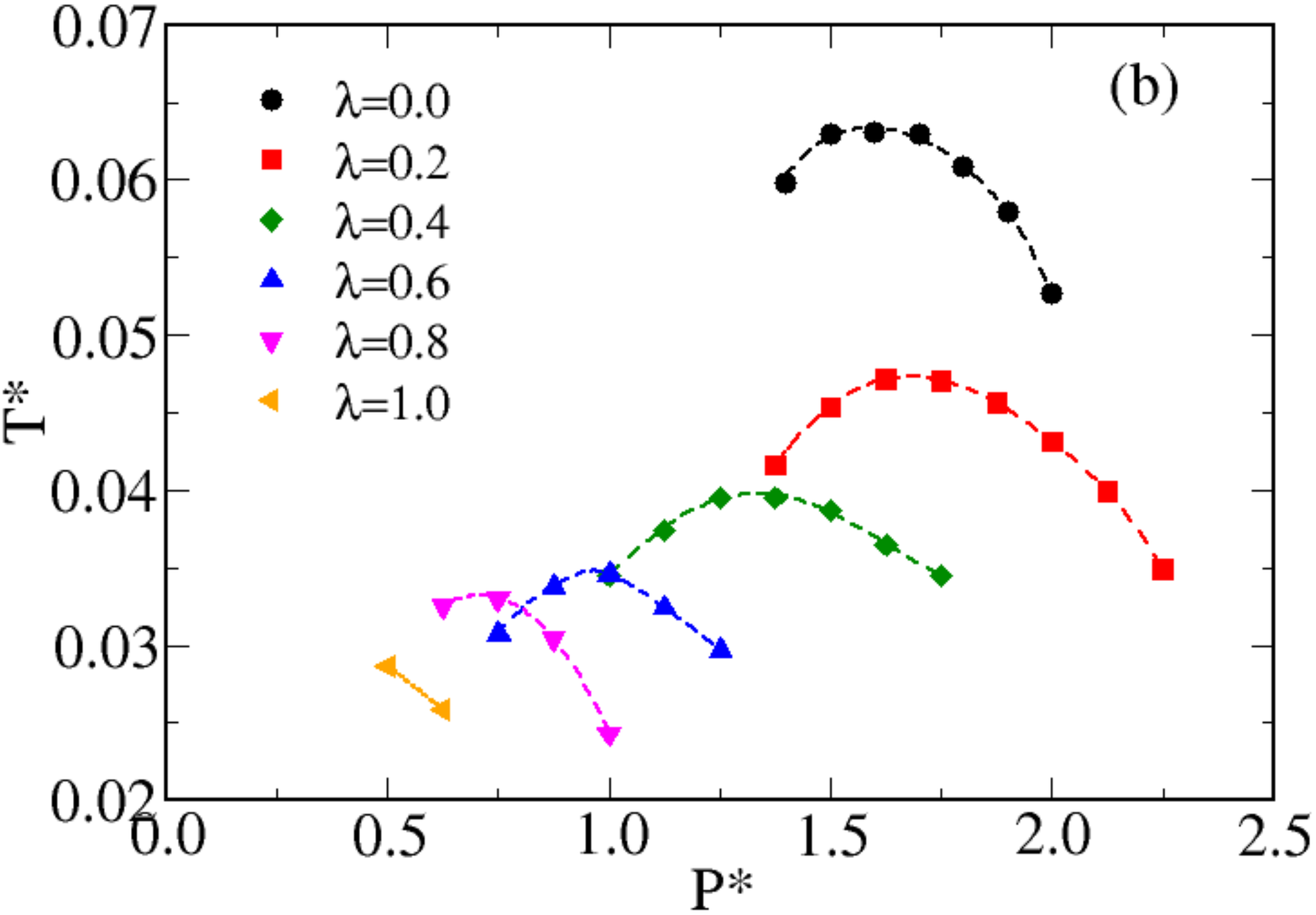} 
\caption{Loci of TMD points at several values of $\lambda$ 
for $\alpha=0.6$ (a) and $\alpha=0.8$ (b) obtained from KH. 
Lines are guides for the eye.}\label{fig:tmd}
\end{center}
\end{figure}
%%%%%%%%%%%%%%%%%%%%%%%%%%%%%%%%%%%%%%%%%%%%%%%%%%%%%%%%%%%%%%%%%%%%%

In a simple fluid, $-S_2$ monotonically increases with 
the density at fixed temperature; conversely, this function shows the 
presence of a maximum and a minimum for systems interacting through CS
potentials~\cite{truskett:00,giaquinta:05}.
Eq.~(\ref{eq:s2}) implies that in principle is not 
possible to calculate $S_2$ within the RISM framework, since $g_{cm}(r)$ is
not available. 
Actually, it is possible to introduce non interacting auxiliary 
sites at the geometric centre of a dimer, with the aim to obtain the pair 
distribution function between the centers of mass of two dimers. However, 
it has been already observed~\cite{Cummings} 
that the site-site radial distribution functions 
show a dependence on these auxiliary sites. Hence, to avoid the introduction 
of spurious effects on $g_{ij}(r)$, we do not 
make use of this prescription.
Also, we have verified that, for $\lambda \leq 0.3$ the
differences between $g_{cm}(r)$ and $g_{ij}(r)$ are not very significant and
an (approximate) theoretical calculation of $S_2$ by putting
$g_{ij}(r)$ in Eq.~(\ref{eq:s2}) is possible.

The behavior of density and structural anomalies for $\alpha=0.6$ 
as functions of $\lambda$ is reported in Fig.~\ref{fig:anom-a06}. The
theory qualitatively follows simulations in predicting the existence of
density and structural anomaly regions, even if they are both underestimated,
as observed for the locus of TMD points for $\lambda=0$ 
(see Fig.~\ref{fig:route}). In this context, in comparison with our previous
simulation study on the MIP dimer fluid~\cite{Munao:16} we have verified 
through more refined calculations that the density anomaly survives till 
to $\lambda=0.1$.  

Density and structural anomalies for $\alpha=0.8$ are investigated in
Fig.~\ref{fig:anom-a08}. As observed for $\alpha=0.6$, the theory correctly 
predicts
the existence of both of them, underestimating the temperatures where they
occur.
Interestingly the agreement between MC and KH for the 
predictions of the TMD progressively
improves upon increasing $\lambda$. This finding suggests that the theory
works better for not too low values of the elongation, where the closeness
of the two interaction sites can affects the accuracy of predictions.  
%%%%%%%%%%%%%%%%%%%%%%%%%%%%%%%%%%%%%%%%%%%%%%%%%%%%%%%%%%%%%%%%%%%%%
\begin{figure}
\begin{center}
\begin{tabular}{c}
\includegraphics[width=8.0cm,angle=0]{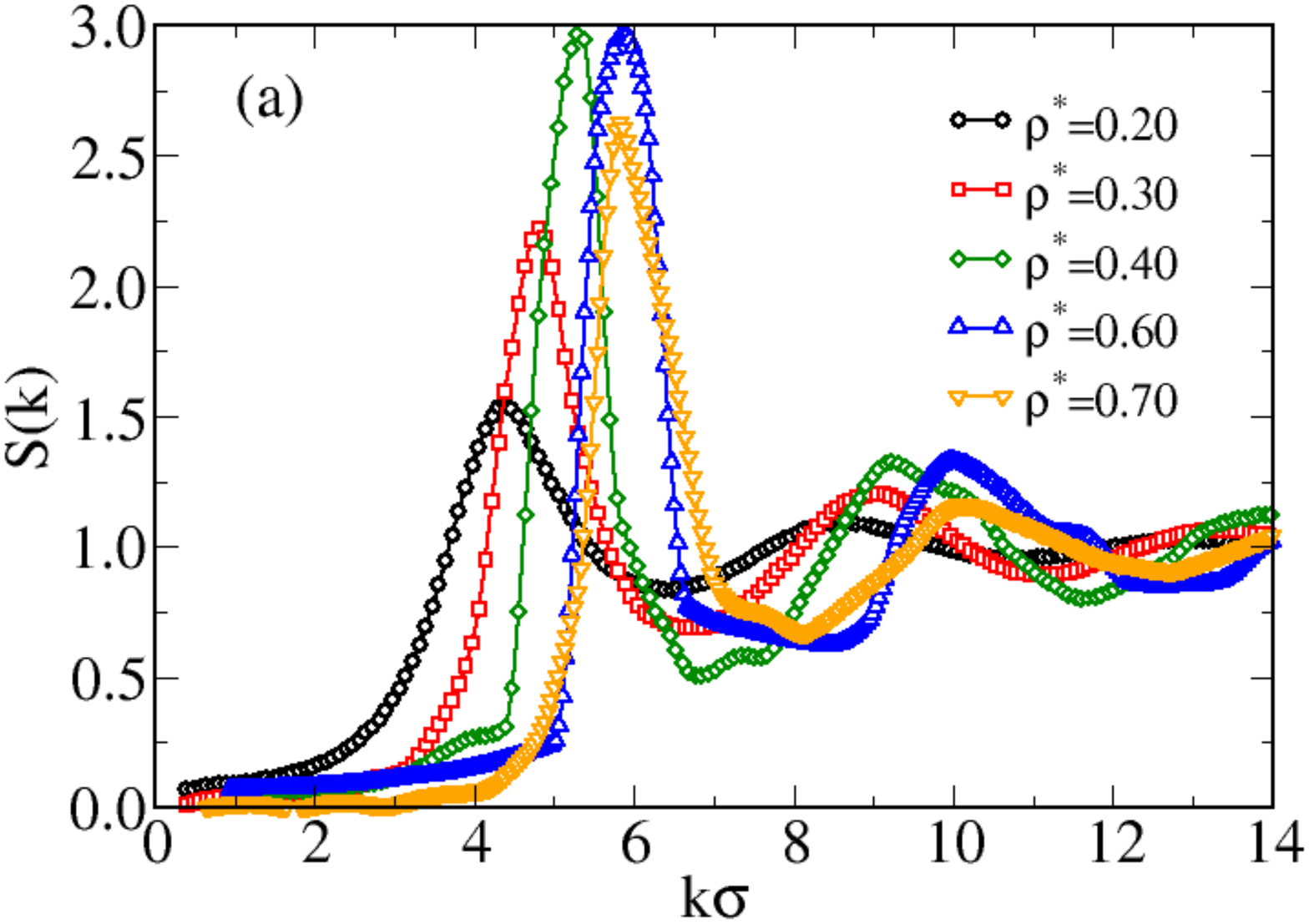} \\
\includegraphics[width=8.0cm,angle=0]{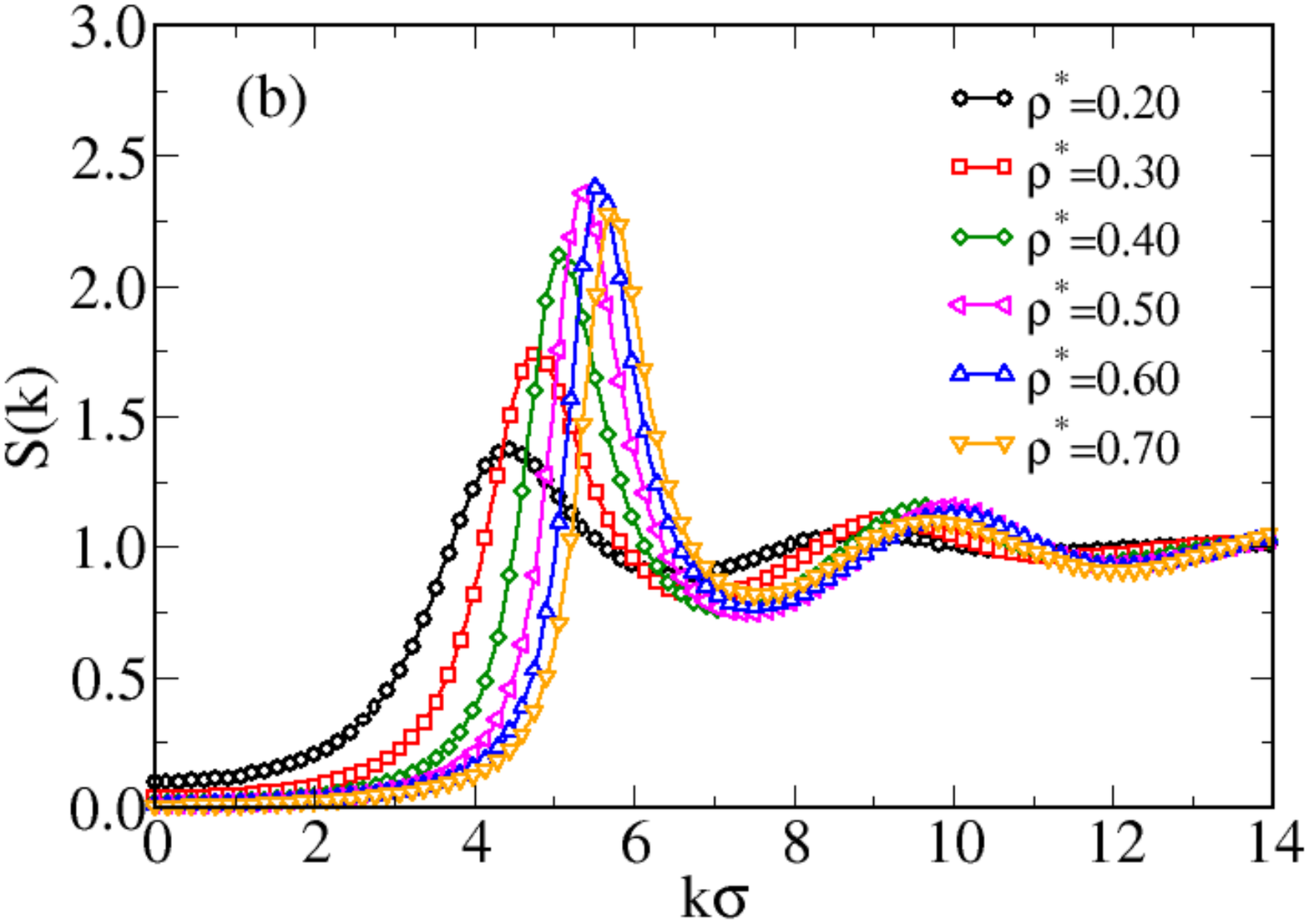} 
\end{tabular}
\caption{Structure factors between the centres of mass of two dimers
obtained from MC simulations (a) and from KH (b) 
for $\lambda=0.2$, $\alpha=0.8$ and $T^*=0.0625$.
}\label{fig:sk}
\end{center}
\end{figure}
%%%%%%%%%%%%%%%%%%%%%%%%%%%%%%%%%%%%%%%%%%%%%%%%%%%%%%%%%%%%%%%%%%%%%

KH predictions for the 
TMD in the whole range of $\lambda$ values are collectively 
reported in Fig~\ref{fig:tmd}: for $\alpha=0.6$ (a) we observe a progressive
shift of the density anomaly regions towards very low temperatures, with a
``saturation'' for $\lambda \geq 0.4$. At the same time, pressure attains
progressively lower values. A similar scenario holds for $\alpha=0.8$ (b),
but in this case the saturation effect is not observed and the density 
anomaly region moves towards lower and lower values of $T^*$. In comparison
with simulations, the theory predicts the existence of the TMD for all
values of $0 \leq \lambda \leq 1$ for both $\alpha=0.6$ and $\alpha=0.8$.
On the other hand, simulations predict the existence of the TMD only if
$\lambda \leq 0.1$ for $\alpha=0.6$ and $\lambda \leq 0.3$ for $\alpha=0.8$.
At higher values of $\lambda$ the development of the density anomaly 
is likely prevented by the onset of solid phases. However, according to KH
predictions, the system persists in a fluid phase down to very low temperatures
and hence the development of the TMD is not hampered by 
a crystallization.  
This observation can be checked by looking at the behavior of
the molecular structure factor $S(k)$ that, for a biatomic
molecule, is equivalent to the structure factor between the centers of mass
of two dimers. According to the qualitative criterion developed by Hansen
and Verlet for the Lennard-Jones potential~\cite{Hansen:69}, 
a fluid system approaches a solid
phase when the height of the first peak of the $S(k)$ exceed 2.85. 
A similar criterion is expected to roughly hold even if a CS potential is 
adopted. In Fig.~\ref{fig:sk} we compare MC and KH $S(k)$ obtained for
$\lambda=0.2$, $\alpha=0.8$ and $T^*=0.625$: simulation data show a progressive
increase of the first peak of the $S(k)$ until it overcomes 3 for densities
between 0.40 and 0.60 and the system becomes solid-like. For density higher
than 0.60, the first peak decreases and the system becomes fluid again.
According to the KH predictions, the increase of the peak is 
equally observed 
but it never becomes greater than 3. Rather, it approaches a maximum at 
2.5 and then decreases. 
%%%%%%%%%%%%%%%%%%%%%%%%%%%%%%%%%%%%%%%%%%%%%%%%%%%%%%%%%%%%%%%%%%%%%
\begin{figure}
\begin{center}
\includegraphics[width=8.0cm,angle=0]{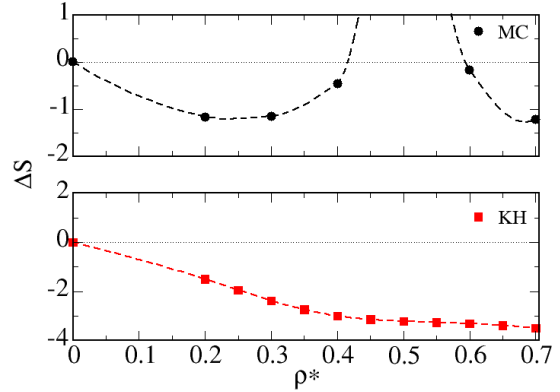} 
\caption{Residual multiparticle entropy $\Delta S$ from
MC simulations (circles) and KH (squares) 
for $\lambda=0.2$, $\alpha=0.8$ and $T^*=0.0625$. Dashed lines are guides
for the eye.}\label{fig:deltas}
\end{center}
\end{figure}
%%%%%%%%%%%%%%%%%%%%%%%%%%%%%%%%%%%%%%%%%%%%%%%%%%%%%%%%%%%%%%%%%%%%%

Finally, we have calculated the 
residual multiparticle entropy $\Delta S$ defined as:
%%%%%%%%%%%%%%%%%%%%%%%%
\begin{equation}
\Delta S= S_{ex}/N - S_2
\end{equation}
%%%%%%%%%%%%%%%%%%%%%%%%
with
%%%%%%%%%%%%%%%%%%%%%%%%
\begin{equation}
S_{ex}/N= (E/N-A_{ex}/N)/T^* 
\end{equation}
%%%%%%%%%%%%%%%%%%%%%%%%
where $A_{ex}/N$ is the excess free energy per particle and
$S_{ex}/N$ is the excess entropy per particle. According to the criterion
developed by Giaquinta and Giunta~\cite{Giunta:92} $\Delta S$ vanishes
whenever a fluid system approaches the solidification. It has been shown 
that the validity of this criterion holds for softly repulsive fluids 
also~\cite{Franz:06}.
In Fig.~\ref{fig:deltas} we report the residual multiparticle entropy
obtained
from MC and KH for $\lambda=0.2$, $\alpha=0.8$ and $T^*=0.0625$.
Simulation data indicate that $\Delta S$ crosses the zero line 
for $0.4 < \rho^* < 0.6$, whereas KH theory never predicts that $\Delta S$
vanishes. Such a result agrees with the behavior of $S(k)$ reported in 
Fig.~\ref{fig:sk}. The emerging picture confirms that KH, while providing
for the reentrant effect typical of anomalous fluids, underestimates the
onset of the solid phase. Summarizing, according to results 
presented in Figs.~\ref{fig:sk} and~\ref{fig:deltas}
for $\lambda=0.2$, $\alpha=0.8$ and $T^*=0.0625$, the system freezes for
$0.4 < \rho^* < 0.6$: on the other hand, the theoretical underestimation
of the onset of the solid phase causes the density anomaly is still observed
in the KH context.   

%%%%%%%%%%%%%%%%%%%%%%%%%%%%%%%%%%%%%%%%%%%%%%%%%%%%%%%%%%%%%%%%%%%%%

\section{Conclusions}

We have investigated density and structural anomalies in 
a dimeric fluid
interacting via a purely repulsive core-softened potential by means of 
integral equation theories. Starting with a
simple monomeric case we have progressively increased the elongation
$\lambda$ till to obtain a tangent configuration. We have also considered two
different conditions, corresponding to one ($\alpha=0.6$) and two 
($\alpha=0.8$) length scales of the intermolecular potential. 
Theoretical results have been systematically compared with already
existing or newly generated Monte Carlo (MC) simulations. We have found
that in the monomeric case the Ornstein-Zernike equation, coupled with
hypernetted chain and Kovalenko-Hirata (KH) closures is able to 
accurately reproduce simulation data for internal energy and 
pressure if the virial equation 
of state is adopted. Nevertheless, the 
loci of density anomalies are 
underestimated by both the closures. Upon increasing $\lambda$ we have
implemented the reference interaction site model (RISM) theory coupled with
KH closure; also, we have adopted a closed form for the calculation 
of pressure, 
available in the context of this closure. The theory is still able to 
reproduce simulation results for pressure, being less predictive for the
internal energy. The fluid structure has been investigated by computing the
site-site radial distribution function $g_{ij}(r)$: MC and KH agree in 
providing a double-peak structure for $\alpha=0.8$, such a feature being
strongly reminiscent of the double length scales of the interaction potential.
This is not the case of $\alpha=0.6$, where only a single peak in $g_{ij}(r)$
is observed. Structural and density anomalies are found for both $\alpha=0.6$
and $\alpha=0.8$: the theory qualitatively agrees 
with simulations in 
predicting their existence, but systematically underestimates the
temperatures where they occur. In comparison with simulations, where the density
anomaly is observed for $\lambda \leq 0.1$ if $\alpha=0.6$ and for 
$\lambda \leq 0.3$ if $\alpha=0.8$,
KH predicts the existence of this anomaly
in the whole range of elongations. This is due to the underestimation of the
solid phase, that in simulations prevents the development of this anomaly for
higher values of $\lambda$. Overall, the theory proves to be reliable in 
providing the existence of structural and thermodynamic anomalies even if
a quantitative prediction is not achieved. We emphasize that this is the 
first application of the RISM approach to investigate these anomalies in a 
fluid interacting through a core-softened potential. The possibility to have
a reliable theoretical tool able to provide a quick estimate of these
anomalies, whereas simulations can become more time demanding, 
set integral equations as promising candidates in describing fluid structure
and phase behaviors of many complex fluids, such as elongated molecules,
polymers and colloidal dimers.

%%%%%%%%%%%%%%%%%%%%%%%%%%%%%%%%%%%%%%%%%%%%%%%%%%%%%%%%%%%%%%%%%%%%%
%\bibliographystyle{iopart-num}
%\bibliography{tmd}

\begin{thebibliography}{10}
\expandafter\ifx\csname url\endcsname\relax
  \def\url#1{{\tt #1}}\fi
\expandafter\ifx\csname urlprefix\endcsname\relax\def\urlprefix{URL }\fi
\providecommand{\eprint}[2][]{\url{#2}}

\bibitem{chandler:72a}
Andersen H~C and Chandler D 1972 {\em J. Chem. Phys.\/} {\bf 57} 1918

\bibitem{chandler:72b}
Chandler D and Andersen H~C 1972 {\em J. Chem. Phys.\/} {\bf 57} 1930

\bibitem{Hansennew}
Hansen J~P and McDonald I~R 2006 {\em Theory of simple liquids, {\rm 3rd
  Ed.}\/} (Academic Press, New York)

\bibitem{chandler:78}
Chandler D 1978 {\em Ann. Rev. Phys. Chem.\/} {\bf 29} 441

\bibitem{lowden:5228}
Lowden L~J and Chandler D 1974 {\em J. Chem. Phys.\/} {\bf 61} 5228

\bibitem{johnson}
Johnson E and Hazoume R~P 1979 {\em J. Chem. Phys.\/} {\bf 70} 1599

\bibitem{lue:5427}
Lue L and Blankschtein D 1995 {\em J. Chem. Phys.\/} {\bf 102} 5427

\bibitem{Kovalenko:02}
Kovalenko A and Hirata F 2002 {\em J. Theor. Comput. Chem.\/} {\bf 1} 381

\bibitem{pettitt:7296}
Pettitt B~M and Rossky P~J 1983 {\em J. Chem. Phys.\/} {\bf 78} 7296

\bibitem{Kvamme:02}
Kvamme B 2002 {\em Phys. Chem. Chem. Phys.\/} {\bf 4} 942

\bibitem{Munao:07}
Costa D, Muna\`o G, Saija F and Caccamo C 2007 {\em J. Chem. Phys.\/} {\bf 127}
  224501

\bibitem{Hirata:03}
Hirata F 2003 {\em Molecular Theory of Solvation\/} (Kluwer Academic,
  Dordrecht)

\bibitem{Munao-cpl}
Muna\`o G, Costa D and Caccamo C 2009 {\em Chem. Phys. Lett.\/} {\bf 470} 240

\bibitem{Munao:PCCP}
Muna\`o G, Costa D, Giacometti A, Caccamo C and Sciortino F 2013 {\em Phys.
  Chem. Chem. Phys.\/} {\bf 15} 20590

\bibitem{Gamez:15}
Muna\`o G, G\'amez F, Costa D, Caccamo C, Sciortino F and Giacometti A 2015
  {\em J. Chem. Phys.\/} {\bf 142} 224904

\bibitem{Munao:11}
Muna\`o G, Costa D, Sciortino F and Caccamo C 2011 {\em J. Chem. Phys.\/} {\bf
  134} 194502

\bibitem{Tripathy-RISM}
Tripathy M and Schweizer K~S 2013 {\em J. Phys. Chem. B\/} {\bf 117} 373

\bibitem{Hirata:13}
Kim B and Hirata F 2013 {\em J. Chem. Phys.\/} {\bf 138} 054108

\bibitem{Hirata:15}
Hirata F and Akasaka K 2015 {\em J. Chem. Phys.\/} {\bf 142} 044110

\bibitem{Sugai:02}
Kinoshita M and Sugay Y 2002 {\em J. Comput. Chem.\/} {\bf 23} 1445

\bibitem{Kobryn:14}
Kobryn A~E, Nikolic D, Lyubimoya O, Gusaroy S and Kovalenko A 2014 {\em J.
  Phys. Chem. B\/} {\bf 118} 12034

\bibitem{Huang:15}
Huang W~J, Blinov N and Kovalenko A 2015 {\em J. Phys. Chem. B\/} {\bf 119}
  5588

\bibitem{Kung:10}
Kung W, Gonz{\'a}lez-Mozuelos P and de~la Cruz M~O 2010 {\em Soft Matter\/}
  {\bf 6} 331

\bibitem{Miyata:08}
Miyata T and Hirata F 2008 {\em J. Comput. Chem.\/} {\bf 29} 871

\bibitem{Miyata:10}
Miyata T, Ikuta Y and Hirata F 2010 {\em J. Chem. Phys.\/} {\bf 133} 044114

\bibitem{Miyata:11}
Miyata T, Ikuta Y and Hirata F 2011 {\em J. Chem. Phys.\/} {\bf 134} 044127

\bibitem{Sato:12}
Kido K, Yokogawa D and Sato H 2012 {\em Chem. Phys. Lett.\/} {\bf 531} 223

\bibitem{debenedetti:96}
Debenedetti P~G 1996 {\em Metastable Liquids. Concepts and Principles\/}
  (Princeton University Press, Princeton)

\bibitem{soper:00}
Soper A~K and Ricci M~A 2000 {\em Phys. Rev. Lett.\/} {\bf 84} 2881

\bibitem{Dokter:05}
Dokter A~M, Woutersen S and Bakker H~J 2005 {\em Phys. Rev. Lett.\/} {\bf 94}
  178301

\bibitem{clark:10}
Clark G~N~I, Hura G~L, Hura J, Soper A~K and Head-Gordon T 2010 {\em Proc. Natl
  Acad. Sci. USA\/} {\bf 107} 14007

\bibitem{Mallamace:13}
Mallamace F, Corsaro C and Stanley H~E 2013 {\em Proc. Natl. Acad. Sci.
  U.S.A.\/} {\bf 110} 4899

\bibitem{Liu:09}
Liu Y, Panagiotopoulos A~Z and Debenedetti P~G 2009 {\em J. Chem. Phys.\/} {\bf
  131} 104508

\bibitem{Poole:11}
Sciortino F, Saika-Voivod I and Poole P~H 2011 {\em Phys. Chem. Chem. Phys.\/}
  {\bf 13} 19759

\bibitem{Franzese:12}
Kesselring T~A, Franzese G, Buldyrev S~V, Hermann H~J and Stanley H~E 2012 {\em
  Scientific Reports\/} {\bf 2} 474

\bibitem{Poole:13}
Poole P~H, Bowles R~K, Saika-Voivod I and Sciortino F 2013 {\em J. Chem.
  Phys.\/} {\bf 138} 034505

\bibitem{palmer:14}
Palmer J~C, Martelli F, Liu Y, Car R, Panagiotopoulous A~Z and Debenedetti P~G
  2014 {\em Nature\/} {\bf 510} 385

\bibitem{smallenburg:15}
Smallenburg F and Sciortino F 2015 {\em Phys. Rev. Lett.\/} {\bf 115} 015701

\bibitem{Stell:70}
Hemmer P~C and Stell G 1970 {\em Phys. Rev. Lett.\/} {\bf 24} 1284

\bibitem{Jagla:99}
Jagla E~A 1999 {\em J. Chem. Phys.\/} {\bf 111} 8980

\bibitem{Malescio:01}
Franzese G, Malescio G, Skibinsky A, Buldyrev S~V and Stanley H~E 2001 {\em
  Nature\/} {\bf 409} 692

\bibitem{Malescio:04}
Skibinsky A, Buldyrev S~V, Franzese G, Malescio G and Stanley H~E 2004 {\em
  Phys. Rev. E\/} {\bf 69} 061206

\bibitem{Wilding:06}
Gibson H~M and Wilding N~B 2006 {\em Phys. Rev. E\/} {\bf 73} 061507

\bibitem{Hus:14}
Hus M and Urbic T 2014 {\em Phys. Rev. E\/} {\bf 90} 062306

\bibitem{Scala:01}
Scala A, Sadr-Lahijany M~R, Giovambattista N, Buldyrev S~V and Stanley H~E 2001
  {\em Phys. Rev. E\/} {\bf 63} 041202

\bibitem{Franz:08}
Malescio G, Saija F and Prestipino S 2008 {\em J. Chem. Phys.\/} {\bf 129}
  241101

\bibitem{Saija:09}
Buldyrev S~V, Malescio G, Angell C~A, Giovambattista N, Prestipino S, Saija F,
  Stanley H~E and Xu L 2009 {\em J. Phys.: Condens. Matter\/} {\bf 21} 504106

\bibitem{Frenkel:09}
Gribova N~V, Fomin Y~D, Frenkel D and Ryzhov V~N 2009 {\em Phys. Rev. E\/} {\bf
  79} 051202

\bibitem{Presti:10}
Prestipino S, Saija F and Malescio G 2010 {\em J. Chem. Phys.\/} {\bf 133}
  144504

\bibitem{Barbosa:10}
de~Oliveira A~B, Neves E~B, Gavazzoni C, Paukowski J~Z, Netz P~A and Barbosa
  M~C 2010 {\em J. Chem. Phys.\/} {\bf 132} 164505

\bibitem{Barbosa:14}
Gavazzoni C, Gonzatti G~K, Pereira L~F, Ramos L~H~C, Netz P~A and Barbosa M~C
  2014 {\em J. Chem. Phys.\/} {\bf 140} 154502

\bibitem{Bordin:16}
Bordin J~R 2016 {\em Physica A\/} {\bf 459} 1

\bibitem{Hirata:05}
Kobryn A~E, Yamaguchi T and Hirata F 2005 {\em J. Mol. Liq.\/} {\bf 119} 7

\bibitem{Urbic:14}
Hu\v{s} M, Muna\`o G and Urbic T 2014 {\em J. Chem. Phys\/} {\bf 141} 164505

\bibitem{Urbic:15}
Muna\`o G and Urbic T 2015 {\em J. Chem. Phys.\/} {\bf 142} 214508

\bibitem{Franz:JPCB}
Malescio G and FSaija 2011 {\em J. Phys. Chem. B\/} {\bf 115} 14091

\bibitem{Franz:MolPhys}
Malescio G, Prestipino S and FSaija 2011 {\em Mol. Phys.\/} {\bf 109} 2837

\bibitem{Munao:16}
Muna\`o G and Saija F 2016 {\em Phys. Chem. Chem. Phys.\/} {\bf 18} 9484

\bibitem{Kovalenko:99}
Kovalenko A and Hirata F 1999 {\em J. Chem. Phys.\/} {\bf 110} 10095

\bibitem{Kovalenko:01}
Kovalenko A and Hirata F 2001 {\em Chem. Phys. Lett.\/} {\bf 349} 496

\bibitem{Ryzhov:03}
Ryzhov V~N and Stishov S~M 2003 {\em Phys. Rev. E\/} {\bf 67} 010201(R)

\bibitem{caccamo}
{Caccamo} C 1996 {\em Phys. Rep.\/} {\bf 274} 1

\bibitem{Morita:60}
Morita T and Hiroike K 1960 {\em Progr. Theor. Phys. (Japan)\/} {\bf 23} 1003

\bibitem{Singer:85}
Singer S~J and Chandler D 1985 {\em Mol. Phys.\/} {\bf 55} 621

\bibitem{Oliveira:06}
de~Oliveira A~B, Netz P~A, Colla T and Barbosa M~C 2006 {\em J. Chem. Phys.\/}
  {\bf 124} 084505

\bibitem{Lomba:07}
Lomba E, Almarza N~G, Martin C and McBride C 2007 {\em J. Chem. Phys.\/} {\bf
  126} 244510

\bibitem{Green}
Nettleton R~E and Green M~S 1958 {\em J. Chem. Phys.\/} {\bf 29} 1365

\bibitem{karplus:96}
Lazaridis T and Karplus M 1996 {\em J. Chem. Phys.\/} {\bf 105} 4294

\bibitem{saitta:03}
Saija F, Saitta A and Giaquinta P~V 2003 {\em J. Chem. Phys.\/} {\bf 119} 3587

\bibitem{esposito:06}
Esposito R, Saija F, Saitta A~M and Giaquinta P~V 2006 {\em Phys. Rev. E\/}
  {\bf 73} 040502

\bibitem{huggins:12}
Huggins D~J 2012 {\em J. Comp. Chem.\/} {\bf 33} 1383

\bibitem{kuffel:14}
Kuffel A, Czapiewski D and Zielkiewicz J 2014 {\em J. Chem. Phys.\/} {\bf 141}
  055103

\bibitem{munao:17}
Muna\`o G, Prestipino S and Saija F {\em {\it in preparation}\/}

\bibitem{truskett:00}
Truskett T~M, Torquato S and Debenedetti P~G 2000 {\em Phys. Rev. E\/} {\bf 62}
  993

\bibitem{giaquinta:05}
Giaquinta P~V and Saija F 2005 {\em ChemPhysChem\/} {\bf 6} 1768

\bibitem{Cummings}
Cummings P~T, Gray C~G and Sullivan D~E 1981 {\em J. Phys. A\/} {\bf 14} 1483

\bibitem{Hansen:69}
Hansen J~P and Verlet L 1969 {\em Phys. Rev.\/} {\bf 184} 151

\bibitem{Giunta:92}
Giaquinta P~V and Giunta G 1992 {\em Physica A\/} {\bf 187} 145

\bibitem{Franz:06}
Saija F, Prestipino S and Giaquinta P~V 2006 {\em J. Chem. Phys.\/} {\bf 124}
  244504

\end{thebibliography}

\providecommand{\newblock}{}

\end{document}